\documentclass[journal]{IEEEtran}
\usepackage{amssymb}
\usepackage{amsmath}
\usepackage{amsfonts}

\newtheorem{theorem}{Theorem}

\newtheorem{definition}[theorem]{Definition}
\newtheorem{example}[theorem]{Example}

\newtheorem{lemma}[theorem]{Lemma}

\newtheorem{proposition}[theorem]{Proposition}
\newtheorem{remark}[theorem]{Remark}

\numberwithin{equation}{subsection}
\numberwithin{theorem}{subsection}
\input{tcilatex}
\begin{document}

\title{The finite harmonic oscillator and its applications to sequences,
communication and radar}
\author{Shamgar Gurevich, Ronny Hadani and Nir Sochen{\Large \ }\thanks{%
S. Gurevich is with the Department of Mathematics, University of California,
Berkeley, CA 94720, USA. Email: shamgar@math.berkeley.edu.} \thanks{%
R. Hadani is with the Department of Mathematics, University of Chicago, IL
60637, USA. Email: hadani@math.uchicago.edu.} \thanks{%
Nir Sochen is with the School of Mathematical Sciences, Tel Aviv University,
Tel Aviv 69978, Israel. Email: sochen@post.tau.ac.il.}\thanks{%
Manuscript received September, 2007.}}
\maketitle

\begin{abstract}
A novel system, called the \textit{oscillator system}, consisting of order
of $p^{3}$ functions (signals) on the finite field $\mathbb{F}_{p},$ with $p$
an odd prime, is described and studied. The new functions are proved to
satisfy good auto-correlation, cross-correlation and low peak-to-average
power ratio properties. Moreover, the oscillator system is closed under the
operation of discrete Fourier transform. Applications of the oscillator
system for discrete radar and digital communication theory are explained.
Finally, an explicit algorithm to construct the oscillator system is
presented.
\end{abstract}


\markboth{IEEE Transactions On Information Theory, Vol. XX, No. Y, Month 2008}
{Gurevich, Hadani and Sochen: Using the Document Class IEEEtran.cls}

\section{Introduction}

\PARstart{O}{ne}-dimensional \textit{analog signals} are complex valued
functions on the real line $%
\mathbb{R}
$. In the same spirit, one-dimensional \textit{digital signals,} also called 
\textit{sequences, }might be considered as complex valued functions on the
finite line $\mathbb{F}_{p}$, i.e., the finite field with $p$ elements. In
both situations the parameter of the line is denoted by $t$ and is referred
to as \textit{time}$.$\textit{\ }In this work, we will consider digital
signals only, which will be simply referred to as signals. The space of
signals $\mathcal{H=}%
\mathbb{C}
(\mathbb{F}_{p})$ is a Hilbert space with the Hermitian product given by 
\begin{equation*}
\left \langle \phi ,\varphi \right \rangle =\tsum \limits_{t\in \mathbb{F}%
p}\phi (t)\overline{\varphi (t)}.
\end{equation*}

A central problem is to construct interesting and useful systems of signals.
Given a\ system $\mathfrak{S}$, there are various desired properties which
appear in the engineering wish list. For example, in various situations \cite%
{GG, HCM} one requires that the signals will be weakly correlated, i.e.,
that for every $\phi \neq \varphi \in \mathfrak{S}$ 
\begin{equation*}
\left \vert \left \langle \phi ,\varphi \right \rangle \right \vert \ll 1.
\end{equation*}%
This property is trivially satisfied if $\mathfrak{S}$ is an orthonormal
basis. Such a system cannot consist of more than $\dim (\mathcal{H)}$
signals, however, for certain applications, e.g., CDMA (Code Division
Multiple Access) \cite{V} a larger number of signals is desired, in that
case the orthogonality condition is relaxed.

During the transmission process, a signal $\varphi $ might be distorted in
various ways. Two basic types of distortions are \emph{time shift }$\varphi
(t)\mapsto \mathsf{L}_{\tau }\varphi (t)=\varphi (t+\tau )$ and \emph{phase
shift }$\varphi (t)\mapsto \mathsf{M}_{w}\varphi (t)=e^{\frac{2\pi i}{p}%
wt}\varphi (t)$, where $\tau ,w\in \mathbb{F}_{p}$. The first type appears
in asynchronous communication and the second type is a Doppler effect due to
relative velocity between the transmitting and receiving antennas. In
conclusion, a general distortion is of the type $\varphi \mapsto \mathsf{M}%
_{w}\mathsf{L}_{\tau }\varphi ,$ suggesting that for every $\varphi \neq
\phi \in \mathfrak{S}$ it is natural to require \cite{HCM} the following
stronger condition 
\begin{equation*}
\left \vert \left \langle \phi ,\mathsf{M}_{w}\mathsf{L}_{\tau }\varphi
\right \rangle \right \vert \ll 1.
\end{equation*}%
Due to technical restrictions in the transmission process, signals are
sometimes required to admit low peak-to-average power ratio \cite{PT}, i.e.,
that for every $\varphi \in \mathfrak{S}$ with $\left \Vert \varphi
\right
\Vert _{2}=1$ 
\begin{equation*}
\max \left \{ \left \vert \varphi (t)\right \vert :t\in \mathbb{F}_{p}\right
\} \ll 1.
\end{equation*}%
Finally, several schemes for digital communication require that the above
properties will continue to hold also if we replace signals from $\mathfrak{S%
}$ by their Fourier transform$.$

In this paper we construct a novel system of (unit) signals $\mathfrak{S}%
_{O} $, consisting of $\ $order of $p^{3}$ signals, where $p$ is an odd
prime, called the \textit{oscillator system}. These signals constitute, in
an appropriate formal sense, a finite analogue for the eigenfunctions of the
harmonic oscillator in the real setting and, in accordance, they share many
of the nice properties of the latter class. In particular, we will prove
that $\mathfrak{S}_{O}$ satisfies the following properties

\begin{enumerate}
\item \emph{Autocorrelation (ambiguity function). }For every $\varphi \in 
\mathfrak{S}_{O}$ we have 
\begin{equation}
\left \vert \left \langle \varphi ,\mathsf{M}_{w}\mathsf{L}_{\tau }\varphi
\right \rangle \right \vert =\left \{ 
\begin{array}{c}
1\text{ \  \  \  \  \  \  \ if \ }\left( \tau ,w\right) =0, \\ 
\leq \frac{2}{\sqrt{p}}\text{ \ if }\left( \tau ,w\right) \neq 0.\text{\ }%
\end{array}%
\right.  \label{cross_eq}
\end{equation}

\item \emph{Crosscorrelation (cross-ambiguity function). }For every $\phi
\neq \varphi \in \mathfrak{S}_{O}$ we have 
\begin{equation}
\left \vert \left \langle \phi ,\mathsf{M}_{w}\mathsf{L}_{\tau }\varphi
\right \rangle \right \vert \leq \frac{4}{\sqrt{p}},\   \label{auto_eq}
\end{equation}%
for every $\tau ,w\in \mathbb{F}_{p}$.

\item \emph{Supremum. }For every signal $\varphi \in \mathfrak{S}_{O}$ we
have%
\begin{equation*}
\max \left \{ \left \vert \varphi (t)\right \vert :t\in \mathbb{F}_{p}\right
\} \leq \frac{2}{\sqrt{p}}.
\end{equation*}

\item \emph{Fourier invariance. }For every signal $\varphi \in \mathfrak{S}%
_{O}$ its Fourier transform $\widehat{\varphi }$ is (up to multiplication by
a unitary scalar) also in $\mathfrak{S}_{O}.\ $
\end{enumerate}

\begin{remark}
Explicit algorithm that generates the oscillator system is given in Appendix
\bigskip \ref{Algorithm}.
\end{remark}

The oscillator system can be extended to a much larger system $\mathfrak{S}%
_{E}$, consisting of order of $p^{5}$ signals if one is willing to
compromise Properties 1 and 2 for a weaker condition. The extended system
consists of all signals of the form $\mathsf{M}_{w}\mathsf{L}_{\tau }\varphi 
$ for $\tau ,w\in \mathbb{F}_{p}$ and $\varphi \in \mathfrak{S}_{O}$. It is
not hard to show that $\# \left( \mathfrak{S}_{E}\right) =$ $p^{2}\cdot \#
\left( \mathfrak{S}_{O}\right) \approx p^{5}$. As a consequence of (\ref%
{cross_eq}) and (\ref{auto_eq}) for every $\varphi \neq \phi \in \mathfrak{S}%
_{E}$ we have 
\begin{equation*}
\left \vert \left \langle \varphi ,\phi \right \rangle \right \vert \leq 
\frac{4}{\sqrt{p}}.
\end{equation*}

The characterization and construction of the oscillator system is
representation theoretic and we devote the rest of the introduction to an
intuitive explanation of the main underlying ideas. As a suggestive model
example we explain first the construction of the well known system of\ chirp
(Heisenberg) signals, deliberately taking a representation theoretic point
of view (see \cite{H2, HCM} for a more comprehensive treatment).

\subsection{Model example (Heisenberg system)}

Let us denote by $\psi :\mathbb{F}_{p}\rightarrow 
\mathbb{C}
^{\times }$ \ the character $\psi (t)=e^{\frac{2\pi i}{p}t}$. We consider
the pair of orthonormal bases $\Delta =\left \{ \delta _{a}:a\in \mathbb{F}%
_{p}\right \} $ and $\Delta ^{\vee }=\left \{ \psi _{a}:a\in \mathbb{F}%
_{p}\right \} $, where $\psi _{a}(t)=\frac{1}{\sqrt{p}}\psi (at)$, and $%
\delta _{a}$ is the Kronecker delta function, $\delta _{a}(t)=1$ if $t=a$
and $\delta _{a}(t)=0$ if $t\neq a.$

\subsubsection{Characterization of the bases $\Delta $ and $\Delta ^{\vee }$}

Let $\mathsf{L}:\mathcal{H\rightarrow H}$ be the time shift operator $%
\mathsf{L}\varphi (t)=\varphi (t+1)$. This operator is unitary and it
induces a homomorphism of groups $\mathsf{L}:\mathbb{F}_{p}\rightarrow U(%
\mathcal{H)}$ given by $\mathsf{L}_{\tau }\varphi (t)=\varphi (t+\tau )$ for
any $\tau \in \mathbb{F}_{p}$.

Elements of the basis $\Delta ^{\vee }$ are character vectors with respect
to the action $\mathsf{L}$, i.e., $\mathsf{L}_{\tau }\psi _{a}=\psi (a\tau
)\psi _{a}$ for any $\tau \in \mathbb{F}_{p}$. In the same fashion, the
basis $\Delta $ consists of character vectors with respect to the
homomorphism $\mathsf{M}:\mathbb{F}_{p}\rightarrow U(\mathcal{H)}$ given by $%
M_{w}\varphi (t)=$ $\psi (wt)\varphi (t)$ for any $w\in \mathbb{F}_{p}$.

\subsubsection{The Heisenberg representation}

The homomorphisms $\mathsf{L}$ and $\mathsf{M}$ can be combined into a
single map $\widetilde{\pi }:\mathbb{F}_{p}\times \mathbb{F}_{p}\rightarrow
U(\mathcal{H)}$ which sends a pair $(\tau ,w)$ to the unitary operator $%
\widetilde{\pi }(\tau ,w)=\psi \left( -\tfrac{1}{2}\tau w\right) \mathsf{M}%
_{w}\circ \mathsf{L}_{\tau }$. The plane $\mathbb{F}_{p}\times \mathbb{F}%
_{p} $ is called the \textit{time-frequency plane} and will be denoted by $V$%
. The map $\widetilde{\pi }$ is not an homomorphism since, in general, the
operators $L_{\tau }$ and $M_{w}$ do not commute. This deficiency can be
corrected if we consider the group $H=V\times \mathbb{F}_{p}$ with
multiplication given by 
\begin{equation*}
(\tau ,w,z)\cdot (\tau ^{\prime },w^{\prime },z^{\prime })=(\tau +\tau
^{\prime },w+w^{\prime },z+z^{\prime }+\tfrac{1}{2}(\tau w^{\prime }-\tau
^{\prime }w)).
\end{equation*}%
The map $\widetilde{\pi }$ extends to a homomorphism $\pi :H\rightarrow U(%
\mathcal{H)}$ given by 
\begin{equation*}
\pi (\tau ,w,z)=\psi \left( -\tfrac{1}{2}\tau w+z\right) \mathsf{M}_{w}\circ 
\mathsf{L}_{\tau }.
\end{equation*}%
The group $H$ is called the \textit{Heisenberg }group and the homomorphism $%
\pi $ is called the \textit{Heisenberg representation}$.$

\subsubsection{Maximal commutative subgroups}

The Heisenberg group is no longer commutative, however, it contains various
commutative subgroups which can be easily described. To every line $L\subset
V$ , that passes through the origin,\ one can associate a maximal
commutative subgroup $A_{L}=\left \{ (l,0)\in V\times \mathbb{F}_{p}:l\in
L\right \} $. It will be convenient to identify the subgroup $A_{L}$ with
the line $L$.

\subsubsection{Bases associated with lines}

Restricting the Heisenberg representation $\pi $ to a subgroup $L$ yields a
decomposition of the Hilbert space $\mathcal{H}$ into a direct sum of
one-dimensional subspaces $\mathcal{H=}\tbigoplus \limits_{\chi }\mathcal{H}%
_{\chi },$ where $\chi $ runs in the set $L^{\vee }$ of (complex valued)
characters of the group $L$. The subspace $\mathcal{H}_{\chi }$ consists of
vectors $\varphi \in \mathcal{H}$ such that $\pi (l)\varphi =\chi (l)\varphi 
$. In other words, the space $\mathcal{H}_{\chi }$ consists of common
eigenvectors with respect to the commutative system of unitary operators $%
\left \{ \pi (l)\right \} _{l\in L}$ such that the operator $\pi \left(
l\right) $ has eigenvalue $\chi \left( l\right) $.

Choosing a unit vector $\varphi _{\chi }\in \mathcal{H}_{\chi \text{ }}$for
every $\chi \in L^{\vee }$ \ we obtain an orthonormal basis $\mathcal{B}%
_{L}=\left \{ \varphi _{\chi }:\chi \in L^{\vee }\right \} $. In particular, 
$\Delta ^{\vee }$ and $\Delta $ are recovered as the bases associated with
the lines $T=\left \{ (\tau ,0):\tau \in \mathbb{F}_{p}\right \} $ and $%
W=\left \{ (0,w):w\in \mathbb{F}_{p}\right \} $ respectively. For a general $%
L$ the signals in $\mathcal{B}_{L}$ are certain kind of chirps. Concluding,
we associated with every line $L\subset V$ \ an orthonormal basis $\mathcal{B%
}_{L},$ and overall we constructed a system of signals consisting of a union
of orthonormal bases 
\begin{equation*}
\mathfrak{S}_{H}\mathfrak{=}\left \{ \varphi \in \mathcal{B}_{L}:L\subset
V\right \} .
\end{equation*}%
For obvious reasons, the system $\mathfrak{S}_{H}$ will be called the 
\textit{Heisenberg }system\textit{. }

\subsubsection{Properties of the Heisenberg system}

\smallskip It will be convenient to introduce the following general notion.
Given two signals $\phi ,\varphi \in \mathcal{H}$, their matrix coefficient
is the function $m_{\phi ,\varphi }:H\rightarrow 
\mathbb{C}
$ \ given by $m_{\phi ,\varphi }(h)=\left \langle \phi ,\pi (h)\varphi
\right \rangle $. In coordinates, if we write $h=\left( \tau ,w,z\right) $
then $m_{\phi ,\varphi }(h)=\psi \left( -\tfrac{1}{2}\tau w+z\right)
\left
\langle \phi ,\mathsf{M}_{w}\circ \mathsf{L}_{\tau }\varphi
\right
\rangle $. When $\phi =\varphi $ the function $m_{\varphi ,\varphi }$
is called the \textit{ambiguity} function of the vector $\varphi $ and is
denoted by $A_{\varphi }=m_{\varphi ,\varphi }$.\smallskip \ 

The system $\mathfrak{S}_{H}$ consists of $p+1$ orthonormal bases\footnote{%
Note that $p+1$ is the number of lines in $V$.}, altogether $p\left(
p+1\right) $ signals and it satisfies the following properties \cite{H2, HCM}

\begin{enumerate}
\item \emph{Autocorrelation}\textbf{. }For every signal $\varphi \in 
\mathcal{B}_{L}$ the function $\left \vert A_{\varphi }\right \vert $ is the
characteristic function of the line $L$, i.e., 
\begin{equation*}
\left \vert A_{\varphi }\left( v\right) \right \vert =\left \{ 
\begin{array}{c}
0,\text{ \ }v\notin L, \\ 
1,\text{\  \ }v\in L.%
\end{array}%
\right.
\end{equation*}

\item \emph{Crosscorrelation}.\textbf{\ }For every $\phi \in \mathcal{B}_{L}$
and $\varphi \in \mathcal{B}_{M}$ where $L\neq M$ we have%
\begin{equation*}
\left \vert m_{\varphi ,\phi }\left( v\right) \right \vert \leq \frac{1}{%
\sqrt{p}},
\end{equation*}%
for every $v\in V$. If $L=M$ then $m_{\varphi ,\phi }$ is the characteristic
function of some translation of the line $L$.

\item \emph{Supremum}\textbf{. }A signal $\varphi \in \mathfrak{S}_{H}$ is a
unimodular function, i.e., $\left \vert \varphi (t)\right \vert =\frac{1}{%
\sqrt{p}}$ for every $t\in \mathbb{F}_{p}$, in particular we have 
\begin{equation*}
\max \left \{ \left \vert \varphi (t)\right \vert :t\in \mathbb{F}_{p}\right
\} =\frac{1}{\sqrt{p}}\ll 1\text{.}
\end{equation*}
\end{enumerate}

\begin{remark}
Note the main differences between the Heisenberg and the oscillator systems.
The oscillator system consists of order of $p^{3}$ signals, while the
Heisenberg system consists of \ order of $p^{2}$ signals. Signals in the
oscillator system admits an ambiguity function concentrated at $0\in V$
(thumbtack pattern) while signals in the Heisenberg system admits ambiguity
function concentrated on a line.
\end{remark}

\subsection{The oscillator system}

Reflecting back on the Heisenberg system we see that each vector $\varphi
\in \mathfrak{S}_{H}$ is characterized in terms of action of the additive
group $G_{a}=\mathbb{F}_{p}$. Roughly, in comparison, each vector in the
oscillator system is characterized in terms of action of the multiplicative
group $G_{m}=\mathbb{F}_{p}^{\times }$. Our next goal is to explain the last
assertion. We begin by giving a model example.

Given a multiplicative character\footnote{%
A multiplicative character is a function $\chi :G_{m}\rightarrow 
\mathbb{C}
$ which satisfies $\chi (xy)=\chi (x)\chi (y)$ for every $x,y\in G_{m}.$} $%
\chi :G_{m}\rightarrow 
\mathbb{C}
^{\times }$, we define a vector $\underline{\chi }\in \mathcal{H}$ by 
\begin{equation*}
\underline{\chi }(t)=\left \{ 
\begin{array}{c}
\frac{1}{\sqrt{p-1}}\chi (t),\text{ \  \  \ }t\neq 0, \\ 
0,\text{ \  \  \  \  \  \  \  \  \  \  \  \  \  \  \ }t=0.%
\end{array}%
\right.
\end{equation*}%
We consider the system $\mathcal{B}_{std}=\left \{ \underline{\chi }:\chi
\in G_{m}^{\vee },\text{ }\chi \neq 1\right \} $, where $G_{m}^{\vee }$ is
the dual group of characters.

\subsubsection{Characterizing the system $\mathcal{B}_{std}$}

For each element $a\in G_{m}$ let $\  \rho _{a}:\mathcal{H\rightarrow H}$ \
be the unitary operator acting by scaling $\rho _{a}\varphi (t)=\varphi (at)$%
. This collection of operators form a homomorphism $\rho :G_{m}\rightarrow U(%
\mathcal{H)}$.

Elements of $\mathcal{B}_{std}$ are character vectors with respect to $\rho $%
, i.e., the vector$\underline{\text{ }\chi }$ satisfies $\rho _{a}\left( 
\underline{\chi }\right) =\chi (a)\underline{\chi }$ for every $a\in G_{m}$.
In more conceptual terms, the action $\rho $ yields a decomposition of the
Hilbert space $\mathcal{H}$ into character spaces $\mathcal{H=}\tbigoplus 
\mathcal{H}_{\chi }$, where $\chi $ runs in $G_{m}^{\vee }$. The system $%
\mathcal{B}_{std}$ consists of a representative unit vector for each space $%
\mathcal{H}_{\chi }$, $\chi \neq 1$.

\subsubsection{The Weil representation}

We would like to generalize the system $\mathcal{B}_{std}$ in a similar
fashion like we generalized the bases $\Delta $ and $\Delta ^{\vee }$ in the
Heisenberg setting. In order to do this we need to introduce several
auxiliary operators.

Let $\rho _{a}:\mathcal{H\rightarrow H}$, $a\in \mathbb{F}_{p}^{\times },$
be the operators acting by $\rho _{a}\varphi (t)=\sigma (a)\varphi (a^{-1}t)$
(scaling), where $\sigma $ is the unique quadratic character of $\mathbb{F}%
_{p}^{\times }$, let $\rho _{T}:\mathcal{H\rightarrow H}$ be the operator
acting by $\rho _{T}\varphi (t)=\psi (t^{2})\varphi (t)$ (quadratic
modulation), and finally let $\rho _{S\text{ }}:\mathcal{H\rightarrow H}$ be
the operator of Fourier transform 
\begin{equation*}
\rho _{S}\varphi (t)=\frac{\nu }{\sqrt{p}}\tsum_{s\in \mathbb{F}_{p}}\psi
(ts)\varphi (s),
\end{equation*}%
where $\nu $ is a normalization constant which will be specified in the body
of the paper. The operators $\rho _{a},\rho _{T}$ and $\rho _{S}$ are
unitary. Let us consider the subgroup of unitary operators generated by $%
\rho _{a},\rho _{S}$ and $\rho _{T}$. This group turns out to be isomorphic
to the finite group $Sp=SL_{2}(\mathbb{F}_{p})$, therefore we obtained a
homomorphism $\rho :Sp\rightarrow U(\mathcal{H)}$. The representation $\rho $
is called the \textit{Weil representation} \cite{W} and it will play a
prominent role in this paper.

\subsubsection{Systems associated with maximal (split) tori}

The group $Sp$ consists of various types of commutative subgroups. We will
be interested in maximal \emph{diagonalizable} commutative subgroups. A
subgroup of this type is called maximal \textit{split} \textit{torus. }The
standard example is the subgroup consisting of all diagonal matrices 
\begin{equation*}
A=\left \{ 
\begin{pmatrix}
a & 0 \\ 
0 & a^{-1}%
\end{pmatrix}%
:a\in G_{m}\right \} ,
\end{equation*}%
which is called the \textit{standard torus}. The restriction of the Weil
representation to a split torus $T\subset Sp$ yields a decomposition of the
Hilbert space $\mathcal{H}$ into a direct sum of character spaces $\mathcal{%
H=}\tbigoplus \mathcal{H}_{\chi }$, where $\chi $ runs in the set of
characters $T^{\vee }$. Choosing a unit vector $\varphi _{\chi }\in \mathcal{%
H}_{\chi \text{ }}$ for every $\chi $ we obtain a collection of orthonormal
vectors $\mathcal{B}_{T}=\left \{ \varphi _{\chi }:\chi \in T^{\vee },\text{ 
}\chi \neq \sigma \right \} $. Overall, we constructed a system 
\begin{equation*}
\mathfrak{S}_{O}^{s}\mathfrak{=}\left \{ \varphi \in \mathcal{B}%
_{T}:T\subset Sp\text{ split}\right \} ,
\end{equation*}%
which will be referred to as the \textit{split oscillator system. We note
that }our initial system $\mathcal{B}_{std}$ \ is recovered as $\mathcal{B}%
_{std}=\mathcal{B}_{A}$.

\subsubsection{Systems associated with maximal (non-split) tori}

>From the point of view of this paper, the most interesting maximal
commutative subgroups in $Sp$ are those which are diagonalizable over an
extension field rather than over the base field $\mathbb{F}_{p}$. A subgroup
of this type is called maximal \textit{non-split torus. }It might be
suggestive to first explain the analogue notion in the more familiar setting
of the field $%
\mathbb{R}
$. \ Here, the standard example of a maximal non-split torus is the circle
group $SO(2)\subset SL_{2}(%
\mathbb{R}
)$. Indeed, it is a maximal commutative subgroup which becomes
diagonalizable when considered over the extension field $%
\mathbb{C}
$ of complex numbers.

The above analogy suggests a way to construct examples of maximal non-split
tori in the finite field setting as well. Let us assume for simplicity that $%
-1$ does not admit a square root in $\mathbb{F}_{p}$. The group $Sp$ acts
naturally on the plane $V=\mathbb{F}_{p}\times \mathbb{F}_{p}$. Consider the
symmetric bilinear form $B$ on $V$ given by 
\begin{equation*}
B((t,w),(t^{\prime },w^{\prime }))=tt^{\prime }+ww^{\prime }.
\end{equation*}

An example of maximal non-split torus is the subgroup $T_{ns}\subset Sp$
consisting of all elements $g\in Sp$ preserving the form $B$, i.e., $g\in
T_{ns}$ if and only if $B(gu,gv)=B(u,v)$ for every $u,v\in V$. In the same
fashion like in the split case, restricting the Weil representation to a
non-split torus $T$ yields a decomposition into character spaces $\mathcal{H=%
}\tbigoplus \mathcal{H}_{\chi }$. Choosing a unit vector $\varphi _{\chi
}\in \mathcal{H}_{\chi }$ for every $\chi \in T^{\vee }$ we obtain an
orthonormal basis $\mathcal{B}_{T}$. Overall, we constructed a system of
signals 
\begin{equation*}
\mathfrak{S}_{O}^{ns}\mathfrak{=}\left \{ \varphi \in \mathcal{B}%
_{T}:T\subset Sp\text{ non-split}\right \} .
\end{equation*}%
The system $\mathfrak{S}_{O}^{ns}$ will be referred to as the \textit{%
non-split oscillator }system\textit{. }The construction of the system $%
\mathfrak{S}_{O}^{ns}$ and the techniques used to study its properties are
the main contribution of this paper.

\subsubsection{Behavior under Fourier transform}

The oscillator system is closed under the operation of Fourier transform,
i.e., for every $\varphi \in \mathfrak{S}_{O}$ we have that (up to
multiplication by a unitary scalar) $\widehat{\varphi }\in \mathfrak{S}_{O}.$
The Fourier transform on the space $%
\mathbb{C}
\left( \mathbb{F}_{p}\right) $ appears as a specific operator $\rho \left( 
\mathrm{w}\right) $ in the Weil representation, where 
\begin{equation*}
\mathrm{w}=%
\begin{pmatrix}
0 & 1 \\ 
-1 & 0%
\end{pmatrix}%
\in Sp.
\end{equation*}%
Given a signal $\varphi \in \mathcal{B}_{T}\subset \mathfrak{S}_{O}$, its
Fourier transform $\widehat{\varphi }=\rho \left( \mathrm{w}\right) \varphi $
is (up to multiplication by a unitary scalar) a signal in $\mathcal{B}%
_{T^{\prime }}$ where $T^{\prime }=\mathrm{w}T\mathrm{w}^{-1}$ . In fact, $%
\mathfrak{S}_{O}$ is closed under all the operators in the Weil
representation! Given an element $g\in Sp$ and a signal $\varphi \in 
\mathcal{B}_{T}$ we have, up to a unitary scalar, that $\rho \left( g\right)
\varphi $ $\in \mathcal{B}_{T^{\prime }}$, where $T^{\prime }=gTg^{-1}$.

In addition, the Weyl element $\mathrm{w}$ is an element in some maximal
torus $T_{\mathrm{w}}$ (the split type of $T_{\mathrm{w}}$ depends on the
characteristic $p$ of the field) and as a result signals $\varphi \in 
\mathcal{B}_{T_{\mathrm{w}}}$ are, in particular, eigenvectors of the
Fourier transform. As a consequences a signal $\varphi \in \mathcal{B}_{T_{%
\mathrm{w}}}$ and its Fourier transform $\widehat{\varphi }$ differ by a
unitary constant, therefore are practically the "same" for all essential
matters.

These properties might be relevant for applications to OFDM (Orthogonal
Frequency Division Multiplexing) \cite{C} where one requires good properties
both from the signal and its Fourier transform.

\subsubsection{Relation to the harmonic oscillator}

Here we give the explanation why functions in the non-split oscillator
system $\mathfrak{S}_{O}^{ns}$ constitute a finite analogue of the
eigenfunctions of the harmonic oscillator in the real setting. The Weil
representation establishes the dictionary between these two, seemingly,
unrelated objects. The argument works as follows.

The one-dimensional harmonic oscillator is given by the differential
operator $D=\partial ^{2}-t^{2}$. The operator $D$ can be exponentiated to
give a unitary representation of the circle group $\rho :SO\left( 2,%
\mathbb{R}
\right) \longrightarrow U\left( \mathcal{H}\right) $ where $\rho \left(
t\right) =e^{itD}$. Eigenfunctions of $D$ are naturally identified with
character vectors with respect to $\rho $. The crucial point is that $\rho $
is the restriction of the Weil representation of $SL_{2}\left( 
\mathbb{R}
\right) $ to the maximal non-split torus $SO\left( 2,%
\mathbb{R}
\right) \subset SL_{2}\left( 
\mathbb{R}
\right) $.

Summarizing, the eigenfunctions of the harmonic oscillator and functions in $%
\mathfrak{S}_{O}^{ns}$ are governed by the same mechanism, namely both are
character vectors with respect to the restriction of the Weil representation
to a maximal non-split torus in $SL_{2}$. The only difference appears to be
the field of definition, which for the harmonic oscillator is the reals and
for the oscillator functions is the finite field.

\subsection{Applications}

Two applications of the oscillator system will be described. The first
application is to the theory of discrete radar. The second application is to
CDMA systems. We will give a brief explanation of these problems, while
emphasizing the relation to the Heisenberg representation.

\subsubsection{Discrete Radar}

The theory of discrete radar is closely related \cite{HCM} to the finite
Heisenberg group $H.$ A radar sends a signal $\varphi (t)$ and obtains an
echo $e(t)$. The goal \cite{Wo} is to reconstruct, in maximal accuracy, the
target range and velocity. The signal $\varphi (t)$ and the echo $e(t)$ are,
principally, related by the transformation%
\begin{equation*}
e(t)=e^{2\pi iwt}\varphi (t+\tau )=\mathsf{M}_{w}\mathsf{L}_{\tau }\varphi
(t),
\end{equation*}%
where the time shift $\tau $ encodes the distance of the target from the
radar and the phase shift encodes the velocity of the target. Equivalently
saying, the transmitted signal $\varphi $ and the received echo $e$ are
related by an action of an element $h_{0}\in H$, i.e., $e=\pi (h_{0})\varphi
.$ The problem of discrete radar can be described as follows. Given a signal 
$\varphi $ and an echo $e=\pi (h_{0})\varphi $ extract the value of $h_{0}$%
\textbf{.} \ 

It is easy to show that $\left \vert m_{\varphi ,e}\left( h\right)
\right
\vert =\left \vert A_{\varphi }\left( h\cdot h_{0}\right)
\right
\vert $ and it obtains its maximum at $h_{0}^{-1}$. This suggests
that a desired signal $\varphi $ for discrete radar should admit an
ambiguity function $A_{\varphi } $ which is highly concentrated around $0\in
H$, which is a property satisfied by signals in the oscillator system
(Property 2).

\begin{remark}
It should be noted that the system $\mathfrak{S}_{O}$ is \ "large"
consisting of $\ $aproximately $p^{3}$ signals. This property becomes
important in a \textit{jamming }scenario.
\end{remark}

\subsubsection{Code Division Multiple Access (CDMA)}

We are considering the following setting.

\begin{itemize}
\item There exists a collection of users $i\in I$, each holding a \textit{bit%
} of information $b_{i}\in 
\mathbb{C}
$ \ (usually $b_{i}$ is taken to be an $N$'th root of unity).

\item Each user transmits his bit of information, say, to a central antenna.
In order to do that, \ he multiplies his bit $b_{i}$ by a private signal $%
\varphi _{i}\in \mathcal{H}$ \ and forms a message $u_{i}=b_{i}\varphi _{i}$.

\item The transmission is carried through a single channel (for example in
the case of cellular communication the channel is the atmosphere), therefore
the message received by the antenna is the sum%
\begin{equation*}
u=\tsum \limits_{i}u_{i}.
\end{equation*}
\end{itemize}

The main problem \cite{V} is to extract the individual bits $b_{i}$ from the
message $u$. The bit $b_{i}$ can be estimated by calculating the inner
product%
\begin{equation}
\left \langle \varphi _{i},u\right \rangle =\tsum \limits_{i}\left \langle
\varphi _{i},u_{j}\right \rangle =\tsum \limits_{j}b_{j}\left \langle
\varphi _{i},\varphi _{j}\right \rangle =b_{i}+\tsum \limits_{j\neq
i}b_{j}\left \langle \varphi _{i},\varphi _{j}\right \rangle .  \notag
\end{equation}%
The last expression above should be considered as a sum of the information
bit $b_{i}$ and an additional noise caused by the interference\textit{\ }of
the other messages. This is the standard scenario also called the \textit{%
Synchronous} scenario. In practice, more complicated scenarios appear, e.g., 
\emph{asynchronous scenario }- in which\emph{\ }each message $u_{i}$ is
allowed to acquire an arbitrary time shift $u_{i}(t)\mapsto u_{i}(t+\tau
_{i})$, \emph{phase shift scenario} - in which each message $u_{i}$ is
allowed to acquire an arbitrary phase shift $u_{i}(t)\mapsto e^{\frac{2\pi i%
}{p}w_{i}t}u_{i}(t)$ and probably also a combination of the two where each
message $u_{i}$ is allowed to acquire an arbitrary distortion of the form $%
u_{i}(t)\mapsto e^{\frac{2\pi i}{p}w_{i}t}u_{i}(t+\tau _{i}).$

The previous discussion suggests that what we are seeking for is a large
system $\mathfrak{S}$ of signals which will enable a reliable extraction of
each bit $b_{i}$ for as many users transmitting through the channel
simultaneously.

\begin{definition}[Stability conditions]
\label{stability_def}\smallskip \ Two unit signals $\phi \neq $ $\varphi $
are called \textbf{stably cross-correlated} if $\left \vert m_{\varphi ,\phi
}\left( v\right) \right \vert \ll 1$ for every $v\in V$. A unit signal $%
\varphi $ is called \textbf{stably auto-correlated\ }if $\left \vert
A_{\varphi }\left( v\right) \right \vert \ll 1$, for every $v\neq 0$. A
system $\mathfrak{S}$ of signals is called a \textbf{stable}\ system if
every signal $\varphi \in \mathfrak{S}$ is stably auto-correlated and any
two different signals $\phi ,\varphi \in \mathfrak{S}$ are stably
cross-correlated.
\end{definition}

Formally what we require for CDMA is a stable system $\mathfrak{S}$. Let us
explain why this corresponds to a reasonable solution to our problem. At a
certain time $t$ the antenna receives a message 
\begin{equation*}
u=\tsum \limits_{i\in J}u_{i},
\end{equation*}%
which is transmitted from a subset of users $J\subset I$. Each message $%
u_{i} $, $i\in J,$ $\ $is of the form $u_{i}=b_{i}e^{\frac{2\pi i}{p}%
w_{i}t}\varphi _{i}(t+\tau _{i})=b_{i}\pi (h_{i})\varphi _{i},$ where $%
h_{i}\in H$. \ In order to extract the bit $b_{i}$ we compute the matrix
coefficient%
\begin{equation*}
m_{\varphi _{i},u}=b_{i}R_{h_{i}}A_{\varphi _{i}}+\#(J-\{i\})o(1),
\end{equation*}%
where $R_{h_{i}}$ is the operator of right translation $R_{h_{i}}A_{\varphi
_{i}}(h)=A_{\varphi _{i}}(hh_{i}).$

If the cardinality of the set $J$ is not too big then by evaluating $%
m_{\varphi _{i},u}$ at $h=h_{i}^{-1}$ we can reconstruct the bit $b_{i}$. It
follows from (\ref{cross_eq}) and (\ref{auto_eq}) that the oscillator system 
$\mathfrak{S}_{O\text{ }}$can support order of $p^{3}$ users, enabling
reliable reconstruction when order of $\sqrt{p}$ users are transmitting
simultaneously.

\subsection{Structure of the paper\textbf{\ }}

Apart from the introduction, the paper consists of three sections and two
appendices. In Section \ref{pre_sec} several basic notions from
representation theory are introduced. Particularly, we define the Heisenberg
and Weil representations over finite fields. In addition, we spend some
space explaining the Weyl transform which is a key tool in our approach to
the Heisenberg and Weil representations. In Section \ref{Grt} the geometric
counterpart of the Weil representation is established, in particular, we
explain the geometric Weyl transform. In Section \ref{osc_sec} we introduce
the oscillator functions and then their main properties are stated in a
series of propositions. Finally, we explain the main ideas in the proof of
each proposition. In Appendix \ref{PT} we give the proofs of all technical
statements which appear in the body of the paper. Finally, in Appendix \ref%
{Algorithm}\ we describe an explicit\ algorithm that generates the
oscillator system $\mathfrak{S}_{O}^{s}$ associated with the collection of
split tori$.$

\subsection{Remark about field extensions\textbf{\ }}

All the results in the introduction were stated for the basic finite field $%
\mathbb{F}_{p},$ where $p$ is an odd prime, for the reason of making the
terminology more accessible. However, in the body of the paper, all the
results are stated and proved for any field extension of the form $\mathbb{F}%
_{q}$ with $q=p^{n}.$

\section{Preliminaries from representation theory\label{pre_sec}}

In this section several fundamental notions from representation theory are
explained. Let $\mathbb{F}_{q}$ denote the finite field consisting of $q$
elements, where $q$ is odd.

\subsection{The Heisenberg group}

Let $(V,\omega )$ be a two-dimensional symplectic vector space over $\mathbb{%
F}_{q}$. Considering $V$ as an abelian group, it admits a non-trivial
central extension 
\begin{equation*}
0\rightarrow \mathbb{F}_{q}\rightarrow H\rightarrow V\rightarrow 0,
\end{equation*}%
called the \textit{Heisenberg }group. \ Concretely, the group $H$ can be
presented as the set $H=V\times \mathbb{F}_{q}$ with the multiplication
given by%
\begin{equation*}
(v,z)\cdot (v^{\prime },z^{\prime })=(v+v^{\prime },z+z^{\prime }+\tfrac{1}{2%
}\omega (v,v^{\prime })).
\end{equation*}%
The center of $H$ is $\ Z=Z(H)=\left \{ (0,z):\text{ }z\in \mathbb{F}%
_{q}\right \} .$ The symplectic group $Sp=Sp(V,\omega )$ acts by
automorphism of $H$ \ through its action on the $V$-coordinate.

\subsection{The Heisenberg representation \label{HR}}

One of the most important attributes of the group $H$ is that it admits,
principally, a unique irreducible representation. The precise statement is
the content of the following celebrated theorem.

\begin{theorem}[Stone-von Neuman]
\label{S-vN}Let $\psi :Z\rightarrow 
\mathbb{C}
^{\times }$ be a non-trivial character of the center. There exists a unique
(up to isomorphism) irreducible unitary representation $(\pi ,H,\mathcal{H)}$
with the center acting by $\psi ,$ i.e., $\pi _{|Z}=\psi \cdot Id_{\mathcal{H%
}}$.
\end{theorem}

The representation $\pi $ which appears in the above theorem will be called
the \textit{Heisenberg representation}.

\subsubsection{Schr\"{o}dinger Models}

The Heisenberg representation admits various different models
(realizations). These models appear in families. In this paper we will be
interested in a specific family associated with \textit{Lagrangian
splittings. }These models are usually referred to in the literature as%
\textit{\ Schr\"{o}dinger models. }Let us explain how these models are
constructed.\textit{\ }

\begin{definition}
A Lagrangian splitting $S$ of $V$ is a pair $(L,M)$ of Lagrangian subspaces%
\textit{\footnote{%
We remind the reader that a Lagrangian subspace $L\subset V$ is maximal
subspace on which the symplectic form vanishes. } such that }$L\cap M=0$.
\end{definition}

Given a Lagrangian splitting $S=(L,M)$ there exists a model $(\pi _{S},H,%
\mathcal{H}_{S})$, where the Hilbert space $\mathcal{H}_{S}$ is $%
\mathbb{C}
(L)$ and the action $\pi _{S}$ is given by the following formulas $\pi
_{S}(l)=\mathsf{L}_{l}$, $\pi _{S}(m)=\mathsf{M}_{\psi (\omega (\cdot ,m))}$
and $\pi _{S}(z)=\mathsf{M}_{\psi (z)}$. Finally, the Hermitian product is
given by $\left \langle f,g\right \rangle =\tsum \limits_{x\in L}f(x)%
\overline{g(x)}$ for $f,g\in \mathcal{H}_{S}$.

\subsection{The Weyl transform\label{Weyl_sub}}

We see from the previous paragraph that the Hilbert space of the Heisenberg
representation can be identified with the Hilbert space of complex valued
functions on $\mathbb{F}_{q}$. This fact has far reaching implications, in
particular, it enables to study properties of functions in representation
theoretic terms. An important tool for doing that is the Weyl transform \cite%
{We2} which is principally equivalent to the operation of taking matrix
coefficient. Given a linear operator $A:\mathcal{H\rightarrow H}$ we can
associate to it a function on the group $H$ defined as follows

\begin{equation*}
W_{A}(h)=\tfrac{1}{\dim \mathcal{H}}Tr(A\pi (h^{-1})).
\end{equation*}%
The transform $W:\mathsf{End}(\mathcal{H)\rightarrow }%
\mathbb{C}
(H)$ is called the \textit{Weyl transform \cite{H1}. }

\subsubsection{Properties of the Weyl transform\label{Weylprop_subsub}}

The image of the Weyl transform is the space $%
\mathbb{C}
(H,\psi ^{-1})$ consisting of functions $f\in 
\mathbb{C}
(H)$ such that $f(z\cdot h)=\psi ^{-1}(z)f(h)$ for every $z\in Z$. Moreover,
it admits a left inverse $\Pi :%
\mathbb{C}
(H)\rightarrow \mathsf{End}(\mathcal{H)}$ given by $\Pi (f)=\tfrac{1}{q}%
\tsum \limits_{h\in H}f(h)\pi (h).$ The transforms $W$ and $\Pi $ are
morphisms of $H\times H$-representations, i.e., if we denote by $\mathsf{%
L,R:H\rightarrow End}\left( 
\mathbb{C}
\left( H\right) \right) $ the left and right regular representations of $H$
then $W_{\pi \left( h_{1}\right) A\pi \left( h_{2}\right)
}=R_{h_{2}^{-1}}L_{h_{1}}W_{A}$. Finally, the transforms $W$ and $\Pi $
exchange composition of operators $\circ $ with group theoretic convolution $%
\ast $, i.e., $W_{A\circ B}=W_{A}\ast W_{B}$ for every $A,B\in \mathsf{End}%
\left( \mathcal{H}\right) $, where we take 
\begin{equation*}
W_{A}\ast W_{B}(h)=\frac{1}{q}\tsum \limits_{h_{1}\cdot
h_{2}=h}W_{A}(h_{1})W_{B}(h_{2}).
\end{equation*}%
It will be somtimes convenient to identify $%
\mathbb{C}
(H,\psi ^{-1})$ with $%
\mathbb{C}
(V)$. Under this identification $\Pi $ is given by $\Pi (f)=\tsum
\limits_{v\in V}f(v)\pi (v)$ and 
\begin{equation}
W_{A}\ast W_{B}(v)=\tsum \limits_{v_{1}+v_{2}=v}\psi \left( -\tfrac{1}{2}%
\omega \left( v_{1},v_{2}\right) \right) W_{A}(v_{1})W_{B}(v_{2}).
\label{tconv_eq}
\end{equation}

\subsubsection{Explicit formulas}

Given a Schr\"{o}dinger model $(\pi _{S},H,\mathcal{H}_{S})$ associated to a
Lagrangian splitting $S=(L,M)$, every operator $A\in \mathsf{End}\left( 
\mathcal{H}_{S}\right) $ can be presented as a function on $L\times L$. In
this presentation, composition is given by convolution of functions $f\circ
g(x,y)=\tsum \limits_{z\in L}f(x,z)g(z,y).$ If we identify $%
\mathbb{C}
(H,\psi ^{-1})$ with $%
\mathbb{C}
(L\times M)$ then the transforms $W$ and $\Pi $ are realized as 
\begin{eqnarray*}
W_{S} &:&%
\mathbb{C}
(L\times L)\rightarrow 
\mathbb{C}
(L\times M), \\
\Pi _{S} &:&%
\mathbb{C}
(L\times M)\rightarrow 
\mathbb{C}
(L\times L),
\end{eqnarray*}%
and are given by $W_{S}=F_{M,L}\circ \alpha ^{\ast }$ and $\Pi _{S}=\beta
^{\ast }\circ F_{M,L}^{-1}$. Here, $\alpha ^{\ast },\beta ^{\ast }$ are
pullbacks via the maps $\alpha ,\beta =\alpha ^{-1}:L\times L\rightarrow
L\times L$ with $\alpha (x,y)=(\tfrac{y-x}{2},\tfrac{x+y}{2})$ and $\beta
(x,y)=\left( y-x,x+y\right) $ and $F_{M,L}:%
\mathbb{C}
(L\times L)\rightarrow 
\mathbb{C}
(L\times M)$ is the Fourier transform along the right $L$-coordinate%
\begin{equation*}
F_{M,L}(f)(l,m)=\tfrac{1}{\dim \mathcal{H}}\tsum \limits_{x\in L}\psi (%
\tfrac{1}{2}\omega (m,x))f(l,x).
\end{equation*}

\subsection{Intertwining maps \label{int_sub}}

Given a pair of Lagrangian splittings $S_{i}=\left( L_{i},M_{i}\right) $, $%
i=1,2$, let us denote by $F_{2,1}=F_{S_{2},S_{1}}$ the composition $\Pi
_{S_{2}}\circ W_{S_{1}}$. The map $F_{2,1}$ is a morphism of $H\times H$%
-representations and will be called \textit{intertwining map. }The map $%
F_{2,1}$ splits into a tensor product $F_{2,1}=F^{L}\boxtimes F^{R}$ where
the specific form of $F^{L}$ and $F^{R}$ depends on the relative position of
the two splittings. We will describe $F^{L}$ and $F^{R}$ explicitly. Let us
denote by $A$ the tautological isomorphism $L_{2}\times M_{2}\overset{\simeq 
}{\longrightarrow }L_{1}\times M_{1}$. The specific form of $F^{L}$ and $%
F^{R}$ depends on the value of $A_{21}$. For every function $f\in 
\mathbb{C}
\left( L_{1}\right) $

\begin{itemize}
\item If $A_{21}\neq 0$ then

\begin{eqnarray*}
F^{L}(f)(x) &=&\tfrac{1}{\dim \mathcal{H}}\tsum \limits_{y\in L_{1}}\psi (%
\tfrac{1}{2}\omega (Dx,x)+\omega (Bx-Cy,y))f(y), \\
F^{R}(f)(x) &=&\tfrac{1}{\dim \mathcal{H}}\tsum \limits_{y\in L_{1}}\psi
(\omega (Cy-Bx,y)-\tfrac{1}{2}\omega (Dx,x))f(y),
\end{eqnarray*}%
where $B=A_{12}-A_{11}A_{21}^{-1}A_{22}$, $C=A_{11}A_{21}^{-1}$ and $%
D=A_{21}^{-1}A_{22}$.

\item If $A_{21}=0$ then \ 
\begin{eqnarray*}
F^{L}(f)(x) &=&\psi (\tfrac{1}{2}\omega (x,A_{11}x))f(A_{11}x), \\
F^{R}(f)(x) &=&\psi (-\tfrac{1}{2}\omega (x,A_{11}x))f(A_{11}x).
\end{eqnarray*}
\end{itemize}

\subsection{The Weil representation \label{Wrep_sub}}

A direct consequence of Theorem \ref{S-vN} is the existence of a projective
representation $\widetilde{\rho }:Sp\rightarrow PGL(\mathcal{H)}$. The
classical construction of $\widetilde{\rho }$ out of the Heisenberg
representation $\pi $ is due to Weil \cite{W}. Considering the Heisenberg
representation $\pi $ and an element $g\in Sp$, one can define a new
representation $\pi ^{g}$ acting on the same Hilbert space via $\pi
^{g}\left( h\right) =\pi \left( g\left( h\right) \right) $. Clearly both $%
\pi $ and $\pi ^{g}$ have central character $\psi $ hence by Theorem \ref%
{S-vN} they are isomorphic. Since the space $\mathsf{Hom}_{H}(\pi ,\pi ^{g})$
is one-dimensional, choosing for every $g\in Sp$ a non-zero representative $%
\widetilde{\rho }(g)\in \mathsf{Hom}_{H}(\pi ,\pi ^{g})$ gives the required
projective representation. In more concrete terms, the projective
representation $\widetilde{\rho }$ is characterized by the formula 
\begin{equation}
\widetilde{\rho }\left( g\right) \pi \left( h\right) \widetilde{\rho }\left(
g^{-1}\right) =\pi \left( g\left( h\right) \right) ,  \label{Eg}
\end{equation}%
for every $g\in Sp$ and $h\in H$. \ It is a peculiar phenomenon of the
finite field setting that the projective representation $\widetilde{\rho }$
\ can be linearized into an honest representation

\begin{theorem}
\label{linearization} There exists a unique\footnote{%
Unique, except in the case the finite field is $\mathbb{F}_{3}$ and $\dim
V=2 $. For the canonical choice in the latter case see \cite{GH1}.} unitary
representation 
\begin{equation*}
\rho :Sp\longrightarrow GL(\mathcal{H)},
\end{equation*}

satisfying the formula (\ref{Eg}).
\end{theorem}

\subsubsection{Weil representation (invariant presentation)}

An elegant description of the Weil representation can be obtained using the
Weyl transform \cite{GH1}. Given an element $g\in Sp$, the operator $\rho
(g) $ can be written as $\rho (g)=\pi (K_{g})$, where $K_{g}$ is the Weyl
transform $K_{g}=W_{\rho (g)}.$ The collection of functions $\{K_{g}\}_{g\in
Sp}$ form a single function $K:Sp\times H\rightarrow 
\mathbb{C}
$. The multiplicativity property of $\rho $ is manifested as 
\begin{equation}
K_{g}\ast K_{h}=K_{gh}\text{ \  \ for every \ }g,h\in Sp.  \label{mul1_eq}
\end{equation}%
These relations can be written as a single relation satisfied by the
function $K$. Consider the maps $m:Sp\times Sp\times V\rightarrow Sp\times V$
and $p_{i}:Sp\times Sp\times V\rightarrow Sp\times V$, $i=1,2$. Here $m$ is
the multiplication map $m(g_{1},g_{2},v)=(g_{1}\cdot g_{2},v)$ and $%
p_{i}(g_{1},g_{2},v)=(g_{i},v)$, $i=1,2$. The multiplicativity relations (%
\ref{mul1_eq}) are equivalent to 
\begin{equation*}
m^{\ast }K=p_{1}^{\ast }K\ast p_{2}^{\ast }K.
\end{equation*}%
Finally, the function $K$ can be explicitly described \cite{GH1} on an
appropriate subset of $Sp$. Let $U\subset Sp$ denote the subset consisting
of all elements $g\in Sp$ such that $g-I$ \ is invertible. For every $g\in U$
and $v\in V$ we have 
\begin{equation}
K(g,v)=\tfrac{1}{\dim \mathcal{H}}\mu (g)\psi (\tfrac{1}{4}\omega (\kappa
(g)v,v)),  \label{kernel_formula}
\end{equation}%
where $\kappa (g)=\frac{g+I}{g-I}$ is the Cayley transform \cite{H1, We1}, $%
\mu (g)=\sigma (-\det (\kappa (g)+I))$ and $\sigma $ is the unique quadratic
character of the multiplicative group $\mathbb{F}_{q}^{\times }$.

\section{Geometric representation theory\label{Grt}}

In this section a geometric counterpart of the Heisenberg and the Weil
representations\ will be established. The approach we employ is called 
\textit{geometrization}, by which sets are replaced by algebraic varieties
(over the finite field) and functions are replaced by $\ell $-adic Weil
sheaves. Informally, algebraic varieties might be thought of as smooth
manifolds and sheaves as vector bundles. Formally, this way of thinking is
far from the true mathematical definition of these "beasts", but still it
gives a good intuitive idea of what is evolving.

\subsection{Preliminaries from algebraic geometry \ }

We denote by $k$ an algebraic closure of the finite field $\mathbb{F}_{q}$.

\subsubsection{Varieties}

\textbf{\ } In this paper, a variety means a smooth quasi projective
algebraic variety over $k$. A variety over $\mathbb{F}_{q}$ is a variety $%
\mathbf{X}$ equipped with an endomorphism $Fr:\mathbf{X\rightarrow X}$
called {\normalsize Frobenius}. We denote by $X$ \ the set of points which
are fixed by Frobenius, i.e., $X=\{x\in \mathbf{X}:Fr(x)=x\}${\normalsize .}%
\medskip

\subsubsection{Sheaves}

We denote by $\mathsf{D}(\mathbf{X)}$ the bounded derived category of
constructible $\ell $-adic sheaves on $\mathbf{X}$ \cite{BBD} and by $%
\mathsf{D}^{p,0}=\mathsf{D}^{p,\geq 0}\cap \mathsf{D}^{p,\leq 0}$ the
Abelian category of perverse sheaves on the variety $\mathbf{X}$. An object $%
\mathcal{F\in }\mathsf{D}^{p,n}$ is called $[n]$-perverse. Note that $%
\mathcal{F}$ is $[n]$-perverse if and only if $\mathcal{F[}n]\in \mathsf{D}%
^{p,0}$, where $[\cdot ]$ denotes the standard cohomological shift functor.
A Weil structure on a sheaf $\mathcal{F\in }\mathsf{D}(\mathbf{X)}$ is an
isomorphism $\theta :\mathcal{F}\overset{\sim }{\longrightarrow }Fr^{\ast }%
\mathcal{F}$. \ A pair $(\mathcal{F},\theta )$ is called a Weil sheaf. By an
abuse of notation we often denote $\theta $ also by $Fr$.

\textbf{Assumption: }We choose once an identification $\overline{%
\mathbb{Q}
}_{\ell }\simeq 
\mathbb{C}
$, hence all sheaves are considered over the complex numbers.

\subsubsection{Sheaf to function correspondence\textbf{\ }}

Given a Weil sheaf $\mathcal{F}$ on $\mathbf{X}$ we can associate to it a
function $f^{\mathcal{F}}:X\rightarrow 
\mathbb{C}
$ by 
\begin{equation*}
f^{\mathcal{F}}(x)=\chi _{Fr}\left( \mathcal{F}_{|x}\right) =\tsum
\limits_{i}(-1)^{i}Tr(Fr_{|H^{i}(\mathcal{F}_{x})}).
\end{equation*}%
This procedure is called \textit{Grothendieck's sheaf-to-function
correspondence} \cite{G, Ga}\textit{. }It interchanges the functors of
pull-back, integration with compact support and tensor product with
pull-back of functions, summation along the fibers and multiplication of
functions respectively.

\subsubsection{Sheaves on one-dimensional varieties}

Let $\mathbf{X}$ be an one-dimensional variety.

\textbf{Elementary sheaves. }An elementary sheaf $\mathcal{F}$ on $\mathbf{X}
$ is an object in $\mathsf{D}(\mathbf{X})$ which is concentrated at a single
degree with no punctual sections \cite{K}. We will denote by $\mathcal{F(}%
t), $ $t\in \mathbf{X,}$ the restriction of $\mathcal{F}$ \ to a punctured
Henselian neighborhood of $t$. \ Alternatively, if we think of $\mathcal{F}$
as a representation of $G=Gal(E/F)$, where $E$ is some separable Galois
extension of the fraction field of $\mathbf{X}$ then $\mathcal{F(}t)$ is the
restriction of $\mathcal{F}$ \ to the inertia subgroup $I_{t}\subset G$.

\textbf{Artin-Schreier sheaf. }$\ $We denote by $\mathcal{L}_{\psi }$ the
Artin-Schreier sheaf \cite{Ga} on the variety $\mathbb{G}_{a}$ which is
associated to an additive character $\psi :\mathbb{F}_{q}\rightarrow 
\mathbb{C}
^{\times }$, in particular we have $f^{\mathcal{L}_{\psi }}=\psi $

\textbf{Kummer sheaf.} We denote by $\mathcal{L}_{\chi }$ the Kummer sheaf
on the variety $\mathbb{G}_{m}$ which is associated to a multiplicative
character $\chi :\mathbb{F}_{q}^{\times }\rightarrow 
\mathbb{C}
^{\times }$, in particular $f^{\mathcal{L}_{\chi }}=\chi .$

\subsection{The geometric Weyl transform}

We use the notations of Subsection \ref{Weyl_sub}. Here we take $(\mathbf{V,}%
\omega )$ to be a dimensional symplectic vector space in the category of
algebraic varieties over $\mathbb{F}_{q}$. Given a Lagrangian splitting $%
\mathbf{S=}\left( \mathbf{L,M}\right) $ of $\mathbf{V}$ we think of the
category $\mathsf{D}(\mathbf{L}\times \mathbf{L})$ as a geometric
counterpart for the vector space of operators $\mathsf{End}\left( \mathcal{H}%
_{S}\right) $. In particular, given a pair of sheaves $\mathcal{F},\mathcal{%
G\in }\mathsf{D}(\mathbf{L}\times \mathbf{L})$ their convolution is defined
by 
\begin{equation}
\mathcal{F\circ G=}\int \limits_{z\in \mathbf{L}}\mathcal{F(}x,z)\otimes 
\mathcal{G}(z,y).  \label{Gconv}
\end{equation}%
where $\int $ denotes the functor of integration with compact support. The
geometric Weyl transform is a functor $W_{S}:\mathsf{D}(\mathbf{L}\times 
\mathbf{L})\rightarrow \mathsf{D}(\mathbf{L}\times \mathbf{M})$ given by $W_{%
\mathbf{S}}=F_{\mathbf{M,L}}\circ \alpha ^{\ast }[2]$. Here $F_{\mathbf{M,L}%
} $ is the $\ell $-adic Fourier transform along the right $\mathbf{L}$%
-coordinate 
\begin{equation*}
F_{\mathbf{M,L}}(\mathcal{F)(}l,m\mathcal{)}\mathcal{=}\int \limits_{x\in 
\mathbf{L}}\mathcal{L}_{\psi }\left( \tfrac{1}{2}\omega (m,x)\right) \otimes 
\mathcal{F(}l,x).
\end{equation*}

\subsubsection{Properties of the geometric Weyl transform}

The functor $W_{\mathbf{S}}$ admits an inverse functor $\Pi _{\mathbf{S}}$,
which is given by $\Pi _{\mathbf{S}}=\beta ^{\ast }\circ F_{\mathbf{M,L}%
}^{-1}$, with $\beta =\alpha ^{-1}$. In addition, the functors $W_{\mathbf{S}%
}$ and $\Pi _{\mathbf{S}}$ interchange between matrix convolution $\circ $
and group theoretic convolution $\ast $, i.e., there exists natural
isomorphisms 
\begin{eqnarray*}
W_{\mathbf{S}}(\mathcal{F\circ G}) &\backsimeq &W_{\mathbf{S}}(\mathcal{%
F)\ast }W_{\mathbf{S}}(\mathcal{G)}, \\
\Pi _{\mathbf{S}}\left( \mathcal{F\ast G}\right) &\simeq &\Pi _{\mathbf{S}}(%
\mathcal{F)\circ }\Pi _{\mathbf{S}}(\mathcal{G)}\text{.}
\end{eqnarray*}%
Here 
\begin{equation*}
W_{\mathbf{S}}(\mathcal{F})\ast W_{\mathbf{S}}(\mathcal{G})(v)=
\end{equation*}%
\begin{equation*}
\int \limits_{v_{1}+v_{2}=v}\mathcal{L}_{\psi }\left( -\tfrac{1}{2}\omega
\left( v_{1},v_{2}\right) \right) \otimes W_{\mathbf{S}}(\mathcal{F}%
)(v_{1})\otimes W_{\mathbf{S}}(\mathcal{G})(v_{2}).
\end{equation*}

Finally, $W_{\mathbf{S}}$ and $\Pi _{\mathbf{S}}$ are compatible with
perverse t-structure, more precisely $W_{\mathbf{S}}$ and $\Pi _{\mathbf{S}}$
shift the perversity degree by $-1$ and $1$ respectively.

\subsection{Intertwining functors}

Given a pair of Lagrangian splittings $\mathbf{S}_{i}=\left( \mathbf{L}_{i},%
\mathbf{M}_{i}\right) $, $i=1,2$, the intertwining functor $F_{\mathbf{S}%
_{2},\mathbf{S}_{1}}$ is the composition of functors $\mathfrak{\Pi }_{%
\mathbf{S}_{2}}\circ W_{\mathbf{S}_{1}}$. The functor $F_{\mathbf{S}_{2},%
\mathbf{S}_{1}}$ establishes an equivalence between the categories $\mathsf{D%
}(\mathbf{L}_{1}\mathbf{\times L}_{1})$ and $\mathsf{D}(\mathbf{L}_{2}%
\mathbf{\times L}_{2})$, it commutes with convolution and sends $\mathsf{D}%
^{p,0}(\mathbf{L}_{1}\mathbf{\times L}_{1})$ to $\mathsf{D}^{p,0}(\mathbf{L}%
_{2}\mathbf{\times L}_{2})$. These properties directly follow from the
properties of the functors $W_{\mathbf{S}}$ and $\Pi _{\mathbf{S}}$.
Finally, we have $F_{\mathbf{S}_{2},\mathbf{S}_{1}}=F^{L}\boxtimes F^{R}$
and:

\begin{itemize}
\item If $A_{21}\neq 0$ then

$F^{L}(\mathcal{G})(x)=$
\end{itemize}

\QTP{Body Math}
$\int \limits_{y\in \mathbf{L}_{1}}\mathcal{L}_{\psi }[\tfrac{1}{2}\omega
(Dx,x)+\omega (Bx,y)+\tfrac{1}{2}\omega (y,Cy)]\otimes \mathcal{G}(y)[1],$

\QTP{Body Math}
$F^{R}(\mathcal{G})(x)=$

\QTP{Body Math}
$\int \limits_{y\in \mathbf{L}_{1}}\mathcal{L}_{\psi }[-\tfrac{1}{2}\omega
(Dx,x)-\omega (Bx,y)-\tfrac{1}{2}\omega (y,Cy)]\otimes \mathcal{G}(y)[1],$

\QTP{Body Math}
with $B,C$ and $D$ given by the same formulas as in subsection \ref{int_sub}.

\begin{itemize}
\item If $A_{21}=0$ then 
\begin{eqnarray*}
F^{L}(\mathcal{G})(x) &=&\mathcal{L}_{\psi }\left( \tfrac{1}{2}\omega
(x,A_{11}x)\right) \otimes \mathcal{G}(A_{11}x), \\
F^{R}(\mathcal{G})(x) &=&\mathcal{L}_{\psi }\left( -\tfrac{1}{2}\omega
(x,A_{11}x)\right) \otimes \mathcal{G}(A_{11}x).
\end{eqnarray*}
\end{itemize}

\subsection{Geometric Weil representation}

We conclude this section by recalling the main result of \cite{GH1}
regarding the existence of a sheaf theoretic counterpart of the Weil
representation. We use the notations from\ Subsection \ref{Wrep_sub}.

\begin{theorem}
\label{GWR}There exists a geometrically irreducible $[\dim \mathbf{Sp}]$%
-perverse Weil sheaf $\mathcal{K}$ of pure weight zero on $\mathbf{Sp\times V%
}$ satisfying the following properties

\begin{enumerate}
\item \textbf{Multiplicativity. }There exists an isomorphism $m^{\ast }%
\mathcal{K\simeq }p_{1}^{\ast }\mathcal{K\ast }p_{2}^{\ast }\mathcal{K}.$

\item \textbf{Function. }We have $f^{\mathcal{K}}=K.$

\item \textbf{Formula}. For every $g\in \mathbf{U}$ we have 
\begin{equation*}
\mathcal{K(}g,v)=\mathcal{L}_{\mu }(g)\otimes \mathcal{L}_{\psi }\left( 
\tfrac{1}{4}\omega (\kappa (g)v,v)\right) [2](1),
\end{equation*}%
where $\mathcal{L}_{\mu }(g)=\mathcal{L}_{\sigma }(-\det (\kappa (g)+I))$.
\end{enumerate}
\end{theorem}

\section{Oscillator functions \label{osc_sec}}

\subsection{The theory of tori\label{tori_sub}}

There exists two conjugacy classes of (rational points of algebraic) tori in 
$Sp\simeq SL_{2}(\mathbb{F}_{q})$. The first system consists of those tori
which are conjugated to the standard diagonal torus%
\begin{equation*}
A=\left \{ 
\begin{pmatrix}
a & 0 \\ 
0 & a^{-1}%
\end{pmatrix}%
:a\in \mathbb{F}_{q}^{\times }\right \} .
\end{equation*}%
A torus in this class is called a \textit{split} torus. The second class
consists of those tori which are not conjugated to $A$. A torus in this
class is called a \textit{non-split }torus (sometimes it is called inert
torus)$.$ All split (non-split) tori are conjugated to one another. The
number of split (non-split) tori is $\# \left( Sp/N_{s}\right) =\tfrac{q(q+1)%
}{2}$ ($\# \left( Sp/N_{ns}\right) =q\left( q-1\right) $), where $N_{s}$ ($%
N_{ns}$) is the normalizer group of some split torus (non-split torus).

Given a torus $T\subset Sp$, the decomposition $\mathcal{H=}\tbigoplus 
\mathcal{H}_{\chi }$ into character spaces depends on the type of $T$. If $T$
\ is a split torus then $\dim \mathcal{H}_{\chi }=1$ unless $\chi =\sigma $,
where $\sigma $ is the unique quadratic character of $T$ (also called 
\textit{Legendre} character), in the latter case $\dim \mathcal{H}_{\sigma
}=2$. If $T$ is a non-split torus then $\dim \mathcal{H}_{\chi }=1$ for
every character $\chi $ which appears in the decomposition, in this case the
quadratic character $\sigma $ does not appear in the decomposition \cite{GH2}%
.

\subsubsection{Geometric projectors}

Below we state the main technical statement of this paper which roughly says
that the character spaces $\mathcal{H}_{\chi }$ can be geometrized.

Given a torus $T\subset Sp$ and a character $\chi \in T^{\vee }$, $\chi \neq
\sigma $, we denote by $P_{\chi }$ the orthogonal projector on the space $%
\mathcal{H}_{\chi }$. Let $W_{P_{\chi }}$ be the Weyl transform of $P_{\chi
} $, we denote by $W_{\chi }$ the normalized function $W_{\chi }=\#(T)\cdot
W_{P_{\chi }}.$

\begin{theorem}
\label{gproj_thm} There exists geometrically irreducible $[1]$-perverse Weil
sheaf $\mathcal{W}_{\chi }$ of pure weight zero on $\mathbf{V}$ such that 
\begin{equation*}
W_{\chi }=f^{\mathcal{W}_{\chi }}.
\end{equation*}
\end{theorem}

For a proof see Appendix \ref{Pgp}.

\subsection{The oscillator system}

Given a torus $T\subset Sp$, choosing for every character $\chi \in T^{\vee
},$ $\chi \neq \sigma $, a unit vector $\varphi _{\chi }\in \mathcal{H}%
_{\chi }$ we obtain a collection of orthonormal vectors $\mathcal{B}%
_{T}=\left \{ \varphi _{\chi }:\chi \neq \sigma \right \} .$ We note, that
when $T$ is non-split, the system $\mathcal{B}_{T}$ is an orthonormal basis.
Considering the union of all these collections, we obtain the oscillator
system

\begin{equation*}
\mathfrak{S}_{O}=\left \{ \varphi \in \mathcal{B}_{T}:T\subset Sp\right \} .
\end{equation*}%
It will be convenient to separate the system $\mathfrak{S}_{O}$ into two
subsystems $\mathfrak{S}_{O}^{s}$ and $\mathfrak{S}_{O}^{ns}$ which
correspond to the split tori and the non-split tori respectively. The
subsystem $\mathfrak{S}_{O}^{s}$ consists of $\tfrac{q(q+1)}{2}$ \
collections, each consisting of $q-2$ orthonormal vectors, altogether $\#%
\mathfrak{S}_{O}^{s}=\frac{q(q+1)(q-2)}{2}$. The non-split subsystem $%
\mathfrak{S}_{O}^{ns}$ consists of $q(q-1)$ collections each consisting of $%
q $ orthonormal vectors, altogether $\# \mathfrak{S}_{O}^{ns}=q^{2}(q-1)$.
The properties of $\mathfrak{S}_{O}$ are summarized in the following
propositions.

\begin{proposition}[Auto-correlations]
\label{auto_prop}For every $\varphi \in \mathcal{B}_{T}$ 
\begin{equation*}
\left \vert A_{\varphi }(h)\right \vert =\left \{ 
\begin{array}{c}
1,\text{ \  \  \  \  \  \  \  \  \ }h\in Z, \\ 
\leq \frac{2}{\sqrt{q}},\text{ \  \ }h\neq Z.%
\end{array}%
\right.
\end{equation*}
\end{proposition}

\begin{proposition}[Cross-correlations]
\label{cross_prop}For every $\varphi \in \mathcal{B}_{T}$ and $\varphi
^{\prime }\in \mathcal{B}_{T^{\prime }}$%
\begin{equation*}
\left \vert m_{\varphi ,\varphi ^{\prime }}(h)\right \vert \leq \frac{4}{%
\sqrt{q}}.
\end{equation*}
\end{proposition}

\begin{proposition}[Supermum]
\label{sup_prop}Let $S=(L,M)$ be a splitting, then for every $\varphi \in 
\mathcal{B}_{T}$ \ 
\begin{equation*}
\sup_{x\in L}\left \vert \varphi (x)\right \vert \leq \frac{2}{\sqrt{q}},%
\text{\ }
\end{equation*}%
where $\varphi $ is realized as a function $\varphi \in \mathcal{H}_{S}=%
\mathbb{C}
(L)$.
\end{proposition}

\begin{remark}
In Proposition \ref{cross_prop}, if $T=T^{\prime },$ $\varphi \neq \varphi
^{\prime },$ then there exists an improved estimate 
\begin{equation*}
\left \vert m_{\varphi ,\varphi ^{\prime }}(h)\right \vert \leq \frac{2}{%
\sqrt{q}}.
\end{equation*}
\end{remark}

In the following subsections we will explain the main arguments in the
proofs of these propositions. The proofs of the technical statements are
given in the appendix.

\subsection{Proof of Proposition \protect \ref{auto_prop}}

Let $T\subset Sp$ be a torus and $\chi \in T^{\vee }$, $\chi \neq \sigma $.
Let $\varphi =\varphi _{\chi }\in \mathcal{H}_{\chi }$ be a unit vector.
Clearly $m_{\varphi ,\varphi }(h)=1$ when $h\in Z$. We would like to show
that $\left \vert m_{\varphi ,\varphi }(h)\right \vert \leq \tfrac{2}{\sqrt{q%
}} $ when $h\notin Z$. In order to do this we will write an explicit
expression for $m_{\varphi ,\varphi }(h)$ and then we will use geometric
techniques to estimate it.

\subsubsection{Explicit expression of the matrix coefficient.\textbf{\  \ }}

Recall $m_{\varphi ,\varphi }(h)=\left \langle \varphi |\pi (h)\varphi
\right \rangle $. Since $\dim \mathcal{H}_{\chi }=1$ we have $\left \langle
\varphi |\pi (h)\varphi \right \rangle =Tr(P_{\chi }\pi (h)),$ which, in
turn, is equal to $\dim \mathcal{H\cdot }W_{P_{\chi }}(h^{-1})$, where $%
P_{\chi }$ is the orthogonal projector on the subspace $\mathcal{H}_{\chi }$%
. The projector $P_{\chi }$ can be written as $P_{\chi }=\frac{1}{\#T}\tsum
\limits_{a\in T}\overline{\chi }(a)\rho (a)$, therefore we can write 
\begin{equation*}
m_{\varphi ,\varphi }(h)=\frac{\dim \mathcal{H}}{\#T}\cdot \tsum
\limits_{a\in T}\overline{\chi }(a)K_{a}\left( h\right) =o\left( 1\right)
\cdot \tsum \limits_{a\in T}\overline{\chi }(a)K_{a}\left( h\right) ,
\end{equation*}%
where we recall that $\dim \mathcal{H}=q$, and $\#T=q\pm 1$, depending on
the type of the torus $T$.

\medskip

\subsubsection{Estimation}

It is enough to estimate $m_{\varphi ,\varphi }(h)$ when $h=v\in V$, $v\neq
0 $.

\begin{proposition}
\label{est_prop}Let $v\in V,$ $v\neq 0,$ then $\left \vert \tsum
\limits_{a\in T}\overline{\chi }(a)K_{a}(v)\right \vert \leq \frac{2}{\sqrt{q%
}}.$
\end{proposition}

As a result we obtain 
\begin{equation*}
\left \vert m_{\varphi ,\varphi }(v)\right \vert =\left \{ 
\begin{array}{c}
1,\text{ \  \  \  \  \  \  \  \  \  \ }v=0, \\ 
\leq c\cdot \frac{2}{\sqrt{q}},\text{ \ }v\neq 0,%
\end{array}%
\right.
\end{equation*}%
where $c=\frac{q}{q-1}$ when $T$ is split and $c=\frac{q}{q+1}$ when $T$ is
non-split.

\subsection{Proof of Proposition \protect \ref{cross_prop}}

Let $T_{i}\subset Sp,$ $i=1,2,$ be a pair of tori and let $\chi _{i}\in
T_{i}^{\vee }$, $\chi _{i}\neq \sigma _{i}$. We choose unit vectors $\varphi
_{i}=\varphi _{\chi _{i}}\in \mathcal{H}_{\chi _{i}}$ $i=1,2,$ and would
like to show that%
\begin{equation*}
\left \vert m_{\varphi _{_{1}},\varphi _{_{2}}}(h)\right \vert \leq \frac{4}{%
\sqrt{q}},\text{ \ }
\end{equation*}%
for every $h\in H$. \ 

Let $P_{i}=P_{\chi _{i}}$ denote the orthogonal projector on $\mathcal{H}%
_{\chi _{i}}$. Our approach will consists of two steps, first we write $%
m_{\varphi _{_{1}},\varphi _{_{2}}}(h)$ in terms of $W_{P_{i}}$ and, second,
we use Theorem \ref{gproj_thm} to obtain an estimate. Explicit calculation
reveals that $\ $%
\begin{equation*}
\left \vert m_{\varphi _{_{1}},\varphi _{_{2}}}(h)\right \vert ^{2}=\dim 
\mathcal{H\cdot }\left \vert W_{P_{_{1}}}\ast Ad_{h}W_{P_{2}}\right \vert
(0),
\end{equation*}%
for every $h\in H$.

Let us denote by $W_{i}$ the normalized function $\#T_{i}\cdot W_{P_{i}}$.

\begin{proposition}
\label{est1_prop} We have $\left \vert W_{1}\ast Ad_{h}W_{2}\right \vert
(0)\leq 16.$
\end{proposition}

Now we can write 
\begin{eqnarray*}
\left \vert m_{\varphi _{_{1}},\varphi _{_{2}}}(h)\right \vert ^{2} &=&\dim 
\mathcal{H\cdot }\left \vert W_{P_{_{1}}}\ast Ad_{h}W_{P_{2}}\right \vert (0)
\\
&\leq &\frac{\dim \mathcal{H}}{\#T_{1}\cdot \#T_{2}}\left \vert W_{1}\ast
Ad_{h}W_{2}\right \vert (0) \\
&\leq &o\left( 1\right) \frac{16}{q},
\end{eqnarray*}

which implies $\left \vert m_{\varphi _{_{1}},\varphi
_{_{2}}}(h)\right
\vert \leq o\left( 1\right) \frac{4}{\sqrt{q}}$.

\subsection{Proof of proposition \protect \ref{est1_prop}}

Let us denote by $C$ the scalar $W_{1}\ast Ad_{h}W_{2}(0)$. Using Theorem %
\ref{gproj_thm} we can describe the scalar $C$ geometrically. Let $\mathcal{W%
}_{i}$ be the sheaf on $\mathbf{V}$ associated to $W_{i}$. We define the
object $\mathcal{C\in }\mathsf{D}\left( \mathbf{pt}\right) $ by 
\begin{equation*}
\mathcal{C}=\left( \mathcal{W}_{1}\ast Ad_{h}\mathcal{W}_{2}\right) _{|0}.
\end{equation*}

The object $\mathcal{C}$ is a Weil object and by the Grothendieck's
Lefschetz trace formula \cite{G} we have $C=f^{\mathcal{C}}$. Since $%
\mathcal{W}_{1}$ and $Ad_{h}\mathcal{W}_{2}$ are of pure weight zero and the
operation of convolution and restriction do not increase weight \cite{D},
this implies that $\mathcal{C}$ is of mixed weight $w(\mathcal{C})\leq 0$.
In more concrete terms, $\mathcal{C}$ is a complex of vector spaces such
that 
\begin{equation*}
\left \vert \mathsf{e.v.}(Fr_{|H^{i}(\mathcal{C})})\right \vert \leq \sqrt{q}%
^{i}.
\end{equation*}

\begin{lemma}[Vanishing lemma]
\label{vanishing_lemma} We have 
\begin{equation*}
\dim H^{i}(\mathcal{C})\leq \left \{ 
\begin{array}{c}
0,\text{ \  \  \  \  \ }i\neq 0, \\ 
16,\text{ \  \  \  \ }i=0.%
\end{array}%
\right.
\end{equation*}
\end{lemma}

Now we can write $\left \vert C\right \vert =\left \vert \chi _{Fr}\left( 
\mathcal{C}\right) \right \vert =\left \vert Tr\left( Fr_{|H^{0}\left( 
\mathcal{C}\right) }\right) \right \vert \leq 16$ which concludes the proof
of the proposition.

\subsubsection{Proof of the vanishing lemma}

The action of $\mathbf{T}_{i}$ on $\mathbf{V}$ yields a decomposition $%
\mathbf{S}_{i}:\mathbf{V=L}_{i}\times \mathbf{M}_{i}$ into eigenspaces$.$
Denote $\Pi _{i}=\Pi _{\mathbf{S}_{i}}$ and $F=F_{\mathbf{S}_{2},\mathbf{S}%
_{1}}$. We have 
\begin{eqnarray*}
\mathcal{C} &\simeq &Tr\left[ \Pi _{2}(\mathcal{W}_{1})\circ \mathsf{Ad}%
_{h}\Pi _{2}(\mathcal{W}_{2})\right] \\
&\simeq &Tr\left[ F\left( \Pi _{1}(\mathcal{W}_{1})\right) \circ Ad_{h}\Pi
_{2}(\mathcal{W}_{2})\right] .
\end{eqnarray*}

Our next goal is to give an explicit description of $\Pi _{i}\left( \mathcal{%
W}_{i}\right) $ as sheaves on $\mathbf{L}_{i}\times \mathbf{L}_{i}$. For
this, we choose vectors $l_{i}\in \mathbf{L}_{i}$ and identify $\tau _{i}:%
\mathbf{T}_{i}\overset{\simeq }{\longrightarrow }\mathbf{L}_{i}$. Denote $%
\mathcal{F}_{\chi _{i}}=\tau _{i!}\left( \mathcal{L}_{\overline{\chi }%
_{i}\sigma }\right) .$

\begin{lemma}
\label{calc_lemma}There exists an isomorphism $\Pi _{i}(\mathcal{W}%
_{i})\simeq \mathcal{F}_{\chi _{i}}\boxtimes \mathcal{F}_{\overline{\chi }%
_{i}}.$
\end{lemma}

Now we can write

$Tr\left[ F\left( \Pi _{1}(\mathcal{W}_{1})\right) \circ Ad_{h}\Pi _{2}(%
\mathcal{W}_{2})\right] \simeq $

\begin{equation*}
\int \limits_{l\in \mathbf{L}_{2}}\left( F^{L}\left( \mathcal{F}_{\chi
_{1}}\right) \otimes \mathcal{F}_{\overline{\chi }_{2}}^{h^{-1}}\right)
[1]\otimes \int \limits_{l\in \mathbf{L}_{2}}\left( F^{R}\left( \mathcal{F}%
_{\chi _{1}}\right) \otimes \mathcal{F}_{\chi _{2}}^{h}\right) [1],
\end{equation*}

where $\mathcal{F}_{\chi _{2}}^{h}$ and $\mathcal{F}_{\overline{\chi }%
_{2}}^{h^{-1}}$ stand for $h\vartriangleright \mathcal{F}_{\chi _{2}}$ and $%
h^{-1}\vartriangleright \mathcal{F}_{\overline{\chi }_{2}}$ respectively.
The result now follows from the following lemma

\begin{lemma}
\label{vanishing1_lemma} We have 
\begin{eqnarray*}
\dim H^{i}\left( \int \limits_{l\in \mathbf{L}_{2}}\left( F^{L}\left( 
\mathcal{F}_{\chi _{1}}\right) \otimes \mathcal{F}_{\overline{\chi }%
_{2}}^{h^{-1}}\right) \right) &\leq &\left \{ 
\begin{array}{c}
4,\text{ \  \  \ }i=1, \\ 
0,\text{ \  \  \ }i=0,%
\end{array}%
\right. \\
\dim H^{i}\left( \int \limits_{l\in \mathbf{L}_{2}}\left( F^{R}\left( 
\mathcal{F}_{\overline{\chi }_{1}}\right) \otimes \mathcal{F}_{\chi
_{2}}^{h}\right) \right) &\leq &\left \{ 
\begin{array}{c}
4,\text{ \  \  \ }i=1, \\ 
0,\text{ \  \  \ }i=0.%
\end{array}%
\right.
\end{eqnarray*}
\end{lemma}

This concludes the proof of the vanishing lemma.

\subsection{Proof of Proposition \protect \ref{sup_prop}}

Let $T\subset Sp$ \ be a torus and $\chi \in T^{\vee }$, $\chi \neq \sigma $%
. We choose a unit vector $\varphi =\varphi _{\chi }\in \mathcal{H}_{\chi }$%
. Let $S=\left( L,M\right) $ be a Lagrangian splitting and $(\pi _{S},H,%
\mathcal{H}_{S})$ be the associated Schr\"{o}dinger model of the Heisenberg
representation. We consider $\varphi $ as a function $\varphi \in \mathcal{H}%
_{S}=%
\mathbb{C}
(L)$ and would like to prove the following estimate 
\begin{equation*}
\left \vert \varphi (x)\right \vert \leq \frac{2}{\sqrt{q}}
\end{equation*}%
for every $x\in L$.

Let us assume that both Lagrangians $L$ and $M$ are not fixed by $T$, the
case when either $L$ or $M$ are fixed by $T$ is easier. Our approach will
consists of two steps, first we interpret the quantity $\varepsilon
=\left
\vert \varphi (x)\right \vert $ in representation theoretic terms
and then we use geometry to obtain an estimate. Recall that we denoted by $%
P_{\chi }$ the orthogonal projector on $\mathcal{H}_{\chi }$, let us denote
by $P_{x}$ the orthogonal projector on the $x$-eigenspace $\mathcal{H}_{x}=%
\mathbb{C}
\delta _{x}\subset \mathcal{H}_{S}$. Explicit calculation reveals that%
\begin{equation*}
\varepsilon ^{2}=Tr(P_{\chi }\cdot P_{x}).
\end{equation*}

It is enough to show that 
\begin{equation*}
\left \vert \varepsilon \right \vert ^{2}\leq \frac{4}{q}.\ 
\end{equation*}

We can write%
\begin{equation*}
Tr(P_{\chi }P_{x})=W_{P_{\chi }}\ast W_{P_{x}}(0).
\end{equation*}

Consider the normalized functions $W_{\chi }=\#T\cdot W_{P_{\chi }}$ and $%
W_{x}=\#L\cdot W_{P_{x}}$. The result follows from the following proposition

\begin{proposition}
\label{est2_prop} We have $\left \vert W_{\chi }\ast W_{x}(0)\right \vert
\leq 4q.$
\end{proposition}

\appendix

\subsection{Proofs of technical statements\label{PT}}

\subsubsection{Proof of Theorem \protect \ref{gproj_thm}\label{Pgp}}

Let $\mathbf{T\subset Sp}$ be the algebraic torus such that $T=\mathbf{T(}%
\mathbb{F}_{q})$. Let $\mathcal{K}$ be the Weil representation sheaf on $%
\mathbf{Sp\times V}$ (Theorem \ref{GWR}). Let us denote by $\mathcal{K}_{%
\mathbf{T}}$ and $\mathcal{K}_{\mathbf{T}^{\times }}$ the restrictions of $%
\mathcal{K}$ to the subvarieties $\mathbf{T\times V}$ and $\mathbf{T}%
^{\times }\mathbf{\times V}$ respectively, where $\mathbf{T}^{\times }$
denotes the punctured torus $\mathbf{T-\{}1\mathbf{\}}$. We define 
\begin{equation*}
\mathcal{W}_{\chi }(v)=\int \limits_{a\in \mathbf{T}}\mathcal{L}_{\overline{%
\chi }}\left( a\right) \otimes \mathcal{K}_{\mathbf{T}}\mathcal{(}a,v).
\end{equation*}

Equivalently, we can write $\mathcal{W}_{\chi }(v)=\pi _{!}(\mathcal{L}%
_{\chi }\otimes \mathcal{K}_{\mathbf{T}}\mathcal{)}$, where \ $\pi :\mathbf{%
T\times V\rightarrow V}$ is the projector on the $\mathbf{V}$-coordinate. By
the Grothendieck's Lefschetz trace formula \cite{G} we have $f^{\mathcal{W}%
_{\chi }}=W_{\chi }$. We would like to show that $\mathcal{W}_{\chi }$ is
geometrically irreducible $[1]$-perverse.

\begin{lemma}
\label{tech1_lemma}The sheaf $\mathcal{K}_{\mathbf{T}}$ is geometrically
irreducible $[\dim \mathbf{T]}$-perverse.
\end{lemma}

Since the functor $\pi _{!}$ is perverse left exact \cite{BBD} hence, using
the previous lemma, we obtain that $\mathcal{W}_{\chi }\in \mathsf{D}%
^{p,\geq 1}$. It is enough to show that $\mathcal{W}_{\chi }\in \mathsf{D}%
^{p,\leq 1}$.

Consider the stratification $\mathbf{V=U}_{0}\cup \mathbf{U}_{1}\cup \mathbf{%
U}_{2}$, where $\mathbf{U}_{2}$ is the open subvariety consisting of all
elements $v\in \mathbf{V}$ which are not eigenvectors with respect to the
action of $\mathbf{T}$, $\mathbf{U}_{1}=\mathbf{V-U}_{2}-\{0\}$ and $\mathbf{%
U}_{0}=\{0\}$.

\begin{lemma}
\label{tech2_lemma} We have $H^{i}\left( \mathcal{W}_{\chi }\right) =0$ for $%
i\neq -1,0$ and 
\begin{eqnarray*}
\dim \text{ }Supp\left( H^{-1}\left( \mathcal{W}_{\chi }\right) \right) &=&2,
\\
\dim \text{ }Supp\left( H^{0}\left( \mathcal{W}_{\chi }\right) \right) &=&0.
\end{eqnarray*}
\end{lemma}

The restrictions on the support of the cohomologies of $\mathcal{W}_{\chi }$
imply that $\mathcal{W}_{\chi }\in \mathsf{D}^{p,\leq 1}$, in fact, it
implies that $\mathcal{W}_{\chi }$ is the middle extension of its
restriction to any open subvariety of $\mathbf{V}$. In particular, $\mathcal{%
W}_{\chi }=j_{!\ast }(\mathcal{W}_{\chi _{|}\mathbf{U}_{2}})$ for $j:\mathbf{%
U}_{2}\hookrightarrow \mathbf{V}$ and because $\mathcal{W}_{\chi _{|\mathbf{U%
}_{2}}}$ is irreducible $[1]$-perverse sheaf, $\mathcal{W}_{\chi }$ is
either. This concludes the proof of the theorem.

\paragraph{Proof of Lemma \protect \ref{tech1_lemma}}

The statement follows from the following two properties of $\mathcal{K}_{%
\mathbf{T}}$. First, the restriction $\mathcal{K}_{\mathbf{T}_{|\mathbf{T}%
^{\times }}}=\mathcal{K}_{\mathbf{T}^{\times }}$ is geometrically
irreducible $[1]$-perverse sheaf, in fact, $\mathcal{K}_{\mathbf{T}^{\times
}}$ is smooth. Second, there exists an isomorphism $m^{\ast }\mathcal{K}_{%
\mathbf{T}}\simeq \mathcal{K}_{\mathbf{T}^{\times }}\ast \mathcal{K}_{%
\mathbf{T}^{\times }}$. Now, consider the map $m:\mathbf{T}^{\times }\times 
\mathbf{T}^{\times }\times \mathbf{V\rightarrow T\times V}$ which is smooth
and surjective. It is enough to show that the pullback $m^{\ast }\mathcal{K}%
_{\mathbf{T}}$ is irreducible $[\dim (\mathbf{T}^{\times }\times \mathbf{T}%
^{\times })]$-perverse. Using the second property we have $m^{\ast }\mathcal{%
K}_{\mathbf{T}}\simeq \mathcal{K}_{\mathbf{T}^{\times }}\ast \mathcal{K}_{%
\mathbf{T}^{\times }}$ where the right hand side is principally an
application of Fourier transform which maintains perversity \cite{KL} so the
statement follows. This concludes the proof of the Lemma.

\paragraph{Proof of Lemma \protect \ref{tech2_lemma}}

We will show that $Supp\left( H^{-1}\left( \mathcal{W}_{\chi }\right)
\right) \subset \mathbf{U}_{2}\cup \mathbf{U}_{1}$ and that $Supp\left(
H^{0}\left( \mathcal{W}_{\chi }\right) \right) \subset \mathbf{U}_{0}$.
First, let $v\in \mathbf{U}_{2}$, we have 
\begin{equation*}
\mathcal{W}_{\chi }(v)=\int \limits_{a\in \mathbf{T}^{\times }}\mathcal{L}_{%
\overline{\chi }}\left( a\right) \otimes \mathcal{L}_{\mu }\left( a\right)
\otimes \mathcal{L}_{\psi }\left( \tfrac{1}{4}\omega (\kappa (a)v,v)\right)
[2](1).
\end{equation*}%
Standard cohomological techniques yields that $\mathcal{W}_{\chi }(v)$ is
concentrated at degree $-1$. In fact, $\mathcal{W}_{\chi _{|\mathbf{U}_{2}}}$
is an irreducible $[1]$-perverse sheaf since it is principally a Fourier
transform of the irreducible perverse sheaf $\mathcal{L}_{\overline{\chi }%
}\otimes \mathcal{L}_{\mu }$. Second, let $v\in \mathbf{U}_{1}$, we have $%
\mathcal{W}_{\chi }(v)=\tint \limits_{\mathbf{T}^{\times }}\mathcal{L}_{%
\overline{\chi }}\otimes \mathcal{L}_{\mu }[2](1)$. Denote $j:\mathbf{T}%
^{\times }\hookrightarrow \mathbf{T}$ and consider the exact triangle of
sheaves on $\mathbf{T}$%
\begin{equation*}
\left( \mathcal{L}_{\overline{\chi }}\otimes \mathcal{L}_{\mu }\right)
_{|1}[-1]\rightarrow j_{!}\left \{ \left( \mathcal{L}_{\overline{\chi }%
}\otimes \mathcal{L}_{\mu }\right) _{|\mathbf{T}^{\times }}\right \}
\rightarrow \mathcal{L}_{\overline{\chi }}\otimes \mathcal{L}_{\mu }.
\end{equation*}

Applying $\pi _{!}[2](1)$ to all the terms in the previous exact sequence we
obtain that $\mathcal{W}_{\chi }(v)=\left( \mathcal{L}_{\chi }\otimes 
\mathcal{L}_{\mu }\right) _{|1}[1](1)$, implying in particular that it is
concentrated at degree $-1$. Finally, let $v\in \mathbf{U}_{0}$, we have $%
\mathcal{W}_{\chi }(0)=\tint \limits_{a\in \mathbf{T}}\mathcal{L}_{\overline{%
\chi }}\left( a\right) \otimes \mathcal{K}_{\mathbf{T}}\mathcal{(}a,0)$.
Using the exact triangle 
\begin{equation*}
j_{!}\left \{ \left( \mathcal{L}_{\overline{\chi }}\otimes \mathcal{K}_{%
\mathbf{T}}\right) _{|\mathbf{T}^{\times }\times 0}\right \} \rightarrow
\left( \mathcal{L}_{\overline{\chi }}\otimes \mathcal{K}_{\mathbf{T}}\right)
_{|\mathbf{T}\times 0}\rightarrow \delta _{1},
\end{equation*}%
we obtain $\pi _{!}\left \{ \left( \mathcal{L}_{\overline{\chi }}\otimes 
\mathcal{K}_{\mathbf{T}}\right) _{|\mathbf{T}^{\times }\times 0}\right \} $
is concentrated at degree $-1,$ $0$.

\subsubsection{Proof of proposition \protect \ref{est_prop}}

Denote $a_{\chi }=\tsum \limits_{a\in T}\overline{\chi }(a)K(a,v)$. Using
formula (\ref{kernel_formula}) we can write 
\begin{equation*}
\tsum \limits_{a\in T}\overline{\chi }(a)K(a,v)=\frac{1}{\dim \mathcal{H}}%
\tsum \limits_{a\in T^{\times }}\overline{\chi }(a)\mu (a)\psi (\tfrac{1}{4}%
\omega (\kappa (a)v,v)),
\end{equation*}%
where $T^{\times }$ denotes the punctured torus $T^{\times }=T-\{1\}$. \ The
last expression can be estimated using standard cohomological techniques. We
have $a_{\chi }=f^{\mathcal{A}_{\chi }}$ where 
\begin{equation*}
\mathcal{A}_{\chi }=\int \limits_{a\in \mathbf{T}^{\times }}\mathcal{L}_{%
\overline{\chi }}\left( a\right) \otimes \mathcal{L}_{\mu }\left( a\right)
\otimes \mathcal{L}_{\psi }(\tfrac{1}{4}\omega (\kappa (a)v,v))(1).
\end{equation*}%
Since integration with compact support does not increase weight \cite{D} $%
\mathcal{A}_{\chi }$ is a Weil object in $\mathsf{D}(\mathbf{pt)}$ of mixed
weight $w(\mathcal{A}_{\chi })\leq 0$. Concretely, this means that $\mathcal{%
A}_{\chi }$ is merely a complex of vector spaces such that $\left \vert 
\mathsf{e.v.}(Fr_{|H^{i}(\mathcal{A}_{\chi })}\right \vert \leq \sqrt{q}^{i}$%
.

\begin{lemma}
\label{tech3_lemma} We have%
\begin{equation*}
\dim H^{i}(\mathcal{A}_{\chi })=\left \{ 
\begin{array}{c}
2,\text{\  \ }i=1, \\ 
0,\text{ \  \ }i\neq 1.%
\end{array}%
\right.
\end{equation*}
\end{lemma}

Now we can write $\left \vert a_{\chi }\right \vert =\left \vert Tr\left(
Fr_{|H^{i}(\mathcal{A}_{\chi })}\right) \right \vert \leq \dim H^{i}(%
\mathcal{A}_{\chi })\cdot \frac{\sqrt{q}}{q}=\frac{2}{\sqrt{q}}.$ This
concludes the proof of the proposition.

\paragraph{Proof of Lemma \protect \ref{tech3_lemma}}

Denote $\mathcal{K}_{\chi }=\mathcal{L}_{\overline{\chi }}\left( a\right)
\otimes \mathcal{L}_{\mu }\left( a\right) \otimes \mathcal{L}_{\psi }\left( 
\tfrac{1}{4}\omega (\kappa (a)v,v)\right) $. Identifying $\  \mathbf{V\simeq }%
\mathbb{A}^{2}$ and $\mathbf{T}^{\times }\mathbf{\simeq }\mathbb{G}%
_{m}^{\times }-\{1\}$ we let $v=(x,y)\neq \left( 0,0\right) $. It is not
hard to verify that the sheaf $\mathcal{L}_{\mu }$, considered as a plain
topological sheaf, is isomorphic to the Kummer sheaf $\mathcal{L}_{\sigma }$
on $\mathbb{G}_{m}^{\times }$ and the sheaf $\mathcal{L}_{\psi }\left( 
\tfrac{1}{4}\omega (\kappa (a)v,v)\right) $ is isomorphic to $\mathcal{L}%
_{\psi }\left( \frac{1}{2}xy\frac{a+1}{a-1}\right) $. We can deduce that $%
\mathcal{K}_{\chi }$ is tame both at $0$ and $\infty $ and it is wild at $1$
with a single break $1$. Since $\mathcal{K}_{\chi }$ is irreducible and
non-constant, the integral $\mathcal{A}_{\chi }=\int \limits_{a\in \mathbf{T}%
^{\times }}\mathcal{K}_{\chi }$ is concentrated at degree $1$, in addition%
\begin{eqnarray*}
\dim H^{1}(\mathcal{A}_{\chi }) &=&Swan_{1}\mathcal{K}_{\chi }-\chi (\mathbb{%
G}_{m}^{\times }) \\
&=&1+1=2.
\end{eqnarray*}%
This concludes the proof of the lemma.

\subsubsection{Proof of Lemma \protect \ref{calc_lemma}}

Fix $i$, denote $\mathbf{T=T}_{i}$, $\chi =\chi _{i}$, $\Pi =\Pi _{i}$ and $%
\mathbf{L}=\mathbf{L}_{i}$, $\mathbf{M=M}_{i}$. \ Since $\mathcal{W}_{\chi }$
is irreducible $[1]$-perverse on $\mathbf{L\times M}$ hence $\Pi (\mathcal{W}%
_{\chi })$ is irreducible $[2]$-perverse on $\mathbf{L}\times \mathbf{L,}$
therefore it is enough to show $\Pi (\mathcal{W}_{\chi })(x,y)\simeq 
\mathcal{F}_{\overline{\chi }}\left( x\right) \boxtimes \mathcal{F}_{\chi
}(y)$ on any open subvariety of $\mathbf{L\times L}$. Let \ $\mathbf{%
U\subset L}\times \mathbf{L}$ denote the open subvariety consisting of $%
(x,y)\in \mathbf{L}\times \mathbf{L}$ so that $x,y\neq 0$ and $x\neq y$. We
have%
\begin{equation*}
\Pi (\mathcal{W}_{\chi })(x,y))=\int \limits_{m\in \mathbf{M}}\mathcal{L}%
_{\psi }\left( \frac{1}{2}\omega \left( x+y,m\right) \right) \otimes 
\mathcal{W}_{\chi }(y-x,m)
\end{equation*}

If we let $\tau :\mathbf{T\rightarrow }GL(\mathbf{L)}$ denote the action of $%
\mathbf{T}$ on $\mathbf{L}$ then explicit computation reveals that

\begin{eqnarray*}
\Pi (\mathcal{W}_{\chi })(x,y) &\simeq &\int \limits_{m\in \mathbf{M}}\int
\limits_{a\in \mathbf{T}^{\times }}\mathcal{L}_{\overline{\chi }\sigma
}(a)\otimes \mathcal{L}_{\psi }\left( \omega \left( \tfrac{\tau (a)y-x}{\tau
(a)-1},m\right) \right) [2] \\
&\simeq &.\int \limits_{a\in \mathbf{T}^{\times }}\mathcal{L}_{\overline{%
\chi }\sigma }(a)\otimes \delta _{\left \{ \tau (a)=\frac{x}{y}\right \}
}[-2][2] \\
&\simeq &\int \limits_{a\in \mathbf{T}^{\times }}\mathcal{L}_{\overline{\chi 
}\sigma }(a)\otimes \delta _{\left \{ \tau (a)=\frac{x}{y}\right \} },
\end{eqnarray*}%
and the last term in is isomorphic to $\mathcal{F}_{\overline{\chi }}\left(
x\right) \boxtimes \mathcal{F}_{\chi }(y)$. This concludes the proof of the
lemma.

\subsubsection{Proof of Lemma \protect \ref{vanishing1_lemma}}

We will prove the second estimate, the first one is proved in exactly the
same manner. Let $h=(v,0)$ and write $v=(l_{2},m_{2})$. First we study $%
\mathcal{F}_{\chi _{2}}^{h}$. We have 
\begin{equation}
\mathcal{F}_{\chi _{2}}^{h}(x)=\mathcal{L}_{\psi }\left( \tfrac{1}{2}\omega
(l_{2},m_{2})+\omega \left( x,m_{2}\right) \right) \otimes \mathcal{F}_{\chi
_{2}}(x+l_{2}).  \notag
\end{equation}

The sheaf $\mathcal{F}_{\chi _{2}}^{h}$ is irreducible $[1]$-perverse,
smooth of rank $1$ on the open subvariety $\mathbf{L}_{2}-\{l_{2}\}$. In
addition, it is tame at $l_{2}$ and wildly ramified at $\infty $ with a
single break equal $1$.

Second, we study $F^{R}\left( \mathcal{F}_{\overline{\chi }_{1}}\right) $.
We assume that we are in the case when $A:$ $\mathbf{L}_{1}\times \mathbf{M}%
_{1}\rightarrow \mathbf{L}_{2}\times \mathbf{M}_{2}$ satisfies $A_{21}\neq 0$%
, the other case is easier and therefore is omitted. We have 
\begin{equation*}
F^{R}(\mathcal{F}_{\overline{\chi }_{1}})(x)=\int \limits_{y\in \mathbf{L}%
_{1}}\mathcal{L}_{\psi }\left( \tfrac{1}{2}\omega (Cy-Bx,y)-\tfrac{1}{2}%
\omega (Dx,x)\right) \otimes \mathcal{F}_{\overline{\chi }_{1}}(y)[1].
\end{equation*}

We assume $C\neq 0$, the analysis when $C=0$ is easier therefore is omitted.
Denote $\mathcal{G}_{1}=\mathcal{L}_{\psi }\left( \tfrac{1}{2}\omega
(Cy,y)\right) \otimes \mathcal{F}_{\overline{\chi }_{1}}$. The sheaf $%
\mathcal{G}_{1}$ is smooth of rank $1$ on $\mathbf{L}_{1}-\{0\}$, it is tame
at $0$, wild at $\infty $ with a single break equal $2$. Denote $\mathcal{G}%
_{2}=$ $\tint \limits_{y\in \mathbf{L}_{1}}\mathcal{L}_{\psi }\left( -\tfrac{%
1}{2}\omega (Bx,y)\right) \otimes \mathcal{G}_{1}(y)[1]$. The sheaf $%
\mathcal{G}_{2}$ is irreducible $[1]$-perverse since it is the (normalized)
Fourier transform of $\mathcal{G}_{1}$. Moreover, for every $x\in \mathbf{L}%
_{2}$, $\mathcal{G}_{2}(x)$ is concentrated at degree $0$ and $\dim \mathcal{%
G}_{2}(x)=Swan_{\infty }\left( \mathcal{G}_{1}\right) =2$, hence $\mathcal{G}%
_{2}$ is smooth of rank 2.

\begin{lemma}
\label{tech4_lemma}We have $\mathcal{G}_{2}(\infty )=\mathcal{G}_{2}(\infty
)_{tame}\oplus \mathcal{G}_{2}(\infty )_{break=2}$, both components are of
dimension 1.
\end{lemma}

Denote $\mathcal{G}_{3}=\mathcal{L}_{\psi }\left( -\tfrac{1}{2}\omega
(Dx,x)\right) \otimes \mathcal{G}_{2}$. The sheaf $\mathcal{G}_{3}$ is
irreducible $[1]$-perverse, smooth of rank $2$ with break decomposition $%
\mathcal{G}_{3}(\infty )=\left( \mathcal{G}_{3}(\infty )\right) _{2}\oplus
\left( \mathcal{G}_{3}(\infty )\right) _{\leq 2}$, both components are of
dimension 1. Finally, denote $\mathcal{G}_{4}=\mathcal{G}_{3}\otimes 
\mathcal{F}_{\chi _{2}}^{h}$. The sheaf $\mathcal{G}_{4}$ is irreducible $%
[1] $-perverse, smooth of rank $2$ on the open subvariety $\mathbf{L}%
_{2}-\{l_{2}\}$, it is tame at $l_{2}$ with break decomposition $\mathcal{G}%
_{4}(\infty )=\mathcal{G}_{4}(\infty )_{2}\oplus \mathcal{G}_{4}(\infty
)_{\leq 2}$, both components are of dimension 1. Now, considering the
integral $\tint \limits_{x\in \mathbf{L}_{2}}\mathcal{G}_{4}$, it is
concentrated at degree $1$ and 
\begin{equation*}
\dim H^{1}\left( \int \limits_{x\in \mathbf{L}_{2}}\mathcal{G}_{4}\right)
=Swan_{\infty }\mathcal{G}_{4}-\chi (\mathbb{G}_{m})\leq 4\text{.}
\end{equation*}

This concludes the proof of the lemma.

\paragraph{Proof of Lemma \protect \ref{tech4_lemma}}

Using the Laumon stationary phase method, the restriction $\mathcal{G}%
_{2}(\infty )$ is a sum of local contributions%
\begin{equation*}
\mathcal{G}_{2}(\infty )=FT_{\psi }loc(0,\infty )\left( \mathcal{G}%
_{1}(0)\right) \oplus FT_{\psi }loc(\infty ,\infty )\left( \mathcal{G}%
_{1}(\infty )\right) .
\end{equation*}%
Here $FT_{\psi }loc(t,\infty )$, $t\in \mathbb{P}^{1}$ denote the Laumon
local Fourier functors. The functors $FT_{\psi }loc(t,\infty ),$ $t=0,\infty 
$ satisfy, in particular, the following properties:

\begin{enumerate}
\item $FT_{\psi }loc(0,\infty )$ sends a tame sheaf of determinant $\mathcal{%
L}_{\chi }$ to a tame sheaf of determinant $\mathcal{L}_{\overline{\chi }}$
of the same rank.

\item $FT_{\psi }loc(\infty ,\infty )$ sends a wild sheaf with a single
break $\frac{a+b}{b}$ of multiplicity $b$ to a wild sheaf with a single
break $\frac{a+b}{a}$ of multiplicity $a.$
\end{enumerate}

Using these two properties we obtain $\mathcal{G}_{2}(\infty )=\left( 
\mathcal{G}_{2}(\infty )\right) _{tame}\oplus \left( \mathcal{G}_{2}(\infty
)\right) _{break=2}$ and $\dim \mathcal{G}_{2}(\infty )_{tame}=\dim \mathcal{%
G}_{2}(\infty )_{break=2}=1$.

\subsubsection{Proof of proposition \protect \ref{est2_prop}}

Let us denote by $C$ the scalar $W_{\chi }\ast W_{x}\left( 0\right) $. We
are going to describe an object $\mathcal{C\in }\mathsf{D}\left( \mathbf{pt}%
\right) $ such that $C=f^{\mathcal{C}}$.

\begin{lemma}
\label{tech5_lemma}There exist geometrically irreducible $[1]$-perverse Weil
sheaf $\mathcal{W}_{x}$ of pure weight $0$ on $\mathbf{V}$ satisfying%
\begin{equation*}
f^{\mathcal{W}_{x}}=W_{x}.
\end{equation*}
\end{lemma}

Denote $\mathcal{C=}\left( \mathcal{W}_{\chi }\ast \mathcal{W}_{x}\right)
_{|0}$. Since convolution does not increase weight \cite{D}, $\mathcal{C}$ \
is a Weil object in $\mathsf{D}(\mathbf{pt)}$ of mixed weight $w(\mathcal{C}%
)\leq 0$. The result now follows from the following statement:

\begin{lemma}
\label{tech6_lemma} We have \ 
\begin{equation*}
\dim H^{i}(\mathcal{C})=\left \{ 
\begin{array}{c}
4,\text{ \  \  \ }i=2, \\ 
0,\text{ \  \  \ }i\neq 2.%
\end{array}%
\right.
\end{equation*}
\end{lemma}

The proof of the proposition now follows easily $C=f^{\mathcal{C}%
}=Tr(Fr_{|H^{2}(\mathcal{C})})\leq 4q.$

\paragraph{Proof of Lemma \protect \ref{tech5_lemma}}

Consider the closed imbedding $i:\mathbf{M}\rightarrow \mathbf{V}$ and
define $\mathcal{W}_{x}=i^{\ast }\mathcal{L}_{\psi (\omega (\cdot ,x))}\left[
2\right] \left( 1\right) $. Clearly, $\mathcal{W}_{x}$ is irreducible $[1]$%
-perverse of pure weight $0$. A direct verification shows that the function $%
W_{x}=f^{\mathcal{W}_{x}}$ satisfies $\Pi \left( W_{x}\right) =P_{x}$.
Concluding the proof of the lemma.

\paragraph{Proof of Lemma \protect \ref{tech6_lemma}}

Let $\mathbf{V=A}\times \mathbf{B}$ be the splitting into eigenspaces of $%
\mathbf{T}$. Denote $\Pi =\Pi _{\mathbf{S}}$. We have 
\begin{equation*}
\mathcal{C}\simeq Tr\left \{ \Pi (\mathcal{W}_{\chi })\circ \Pi (\mathcal{W}%
_{x})\right \} .
\end{equation*}

Both $\Pi (\mathcal{W}_{\chi })$ and $\Pi (\mathcal{W}_{x})$ are irreducible 
$[2]$-perverse and can be calculated explicitly. We know $\Pi (\mathcal{W}%
_{\chi })\simeq \mathcal{F}_{\chi }\boxtimes \mathcal{F}_{\overline{\chi }}$
(Lemma \ref{calc_lemma}). We have%
\begin{equation*}
\Pi (\mathcal{W}_{x})(a_{1},a_{2})\simeq \int \limits_{b\in \mathbf{B}}%
\mathcal{L}_{\psi }\left( \tfrac{1}{2}\omega (a_{1}+a_{2},b)\right) \otimes 
\mathcal{W}_{x}(a_{2}-a_{1},b).
\end{equation*}

Let $R:\mathbf{A\rightarrow B}$ be the linear map characterized by the
property that $\omega (a,x)=\omega (R(a),x)$ for every $a\in \mathbf{A}$. We
obtain 
\begin{eqnarray*}
\Pi (\mathcal{W}_{x})(a_{1},a_{2}) &\simeq &\mathcal{L}_{\psi }\left( \tfrac{%
1}{2}\omega (R(a_{2}-a_{1}),a_{1}+a_{2})\right) \\
&&\otimes \mathcal{L}_{\psi }\left( \omega (\left( I-R\right)
(a_{2}-a_{1}),x)\right) \\
&\simeq &\mathcal{F}_{x}(a_{1})\boxtimes \overline{\mathcal{F}_{x}}(a_{2}),
\end{eqnarray*}

where 
\begin{eqnarray*}
\mathcal{F}_{x}(a_{2}) &\simeq &\mathcal{L}_{\psi }\left( \tfrac{1}{2}\omega
\left( R(a_{2}),a_{2}\right) +\omega \left( \left( I-R\right)
(a_{2}),x\right) \right) , \\
\overline{\mathcal{F}_{x}}\left( a_{1}\right) &\simeq &\mathcal{L}_{\psi
}\left( -\tfrac{1}{2}\omega \left( R(a_{1}),a_{1}\right) -\omega \left(
\left( I-R\right) (a_{1}),x\right) \right) .
\end{eqnarray*}

Therefore we can write 
\begin{eqnarray*}
\mathcal{C} &\simeq &Tr\left \{ \left( \mathcal{F}_{\chi }\boxtimes \mathcal{%
F}_{\overline{\chi }}\right) \circ (\mathcal{F}_{x}\boxtimes \overline{%
\mathcal{F}}_{x})\right \} \\
&\simeq &\left( \int \limits_{\mathbf{L}}\mathcal{F}_{\chi }\otimes 
\overline{\mathcal{F}}_{x}\right) \otimes \left( \int \limits_{\mathbf{L}}%
\mathcal{F}_{\overline{\chi }}\otimes \mathcal{F}_{x}\right) .
\end{eqnarray*}

The statement now follows from

\begin{lemma}
\label{tech8_lemma} We have 
\begin{eqnarray*}
\dim H^{i}\left( \int \limits_{\mathbf{L}}\mathcal{F}_{\chi }\otimes 
\overline{\mathcal{F}}_{x}\right) &=&\left \{ 
\begin{array}{c}
2,\text{ \ }i=1, \\ 
0,\text{ \ }i\neq 1,%
\end{array}%
\right. \\
\dim H^{i}\left( \int \limits_{\mathbf{L}}\mathcal{F}_{\overline{\chi }%
}\otimes \mathcal{F}_{x}\right) &=&\left \{ 
\begin{array}{c}
2,\text{ \ }i=1, \\ 
0,\text{ \ }i\neq 1.%
\end{array}%
\right.
\end{eqnarray*}
\end{lemma}

\paragraph{Proof of Lemma \protect \ref{tech8_lemma}}

The sheaf $\mathcal{F}_{\chi }$ is irreducible perverse, smooth of rank $1$
on $\mathbf{L}-\{0\}$, tame at $0$ and $\infty $. The sheaf $\overline{%
\mathcal{F}}_{x}$ is irreducible perverse, smooth of rank $1$, wild at $%
\infty $ with a single break equal $2$. \ Therefore, the sheaf $\mathcal{G=F}%
_{\chi }\otimes \overline{\mathcal{F}}_{x}$ is irreducible perverse, smooth
of rank $1$ on $\mathbf{L}-\{0\}$, tame at $0$, wild at $\infty $ with a
single break equal $2$. The integral $\tint \limits_{\mathbf{L}}\mathcal{G}$
is concentrated at cohomological degree $1$ and $\dim H^{1}\left( \tint
\limits_{\mathbf{L}}\mathcal{G}\right) =Swan_{\infty }\mathcal{G}-\chi
\left( \mathbb{G}_{m}\right) =2$ . The second estimate is proved in the same
manner. This concludes the proof of the lemma.

\subsection{\textbf{Construction of the oscillator system\label{Algorithm}}}

\subsubsection{Algorithm}

We describe an explicit\ algorithm that generates the oscillator system $%
\mathfrak{S}_{O}^{s}$ associated with the collection of split tori in $Sp.$

\paragraph{Tori}

Consider the standard diagonal torus 
\begin{equation*}
A=\left \{ 
\begin{pmatrix}
a & 0 \\ 
0 & a^{-1}%
\end{pmatrix}%
;\text{ }a\in \mathbb{F}_{p}^{\times }\right \} .
\end{equation*}

Every split torus in $Sp$ is conjugated to the torus $A$, which means that
the collection $\mathcal{T}$ of all split tori in $Sp$ can be written as 
\begin{equation*}
\mathcal{T}=\{gAg^{-1};\ g\in Sp\}.
\end{equation*}

\paragraph{Parametrization}

A direct calculation reveals that every torus in $\mathcal{T}$ can be
written as $gAg^{-1}$ for an element $g$ of the form 
\begin{equation}
g=%
\begin{pmatrix}
1 & b \\ 
c & 1+bc%
\end{pmatrix}%
,\text{ }b,c\in \mathbb{F}_{p}.  \label{form_eq}
\end{equation}

If $b=0$, this presentation is unique: In the case $b\neq 0$, an element $%
\widetilde{g}$ represents the same torus as $g$ if and only if \ it is of
the form 
\begin{equation*}
\widetilde{g}=%
\begin{pmatrix}
1 & b \\ 
c & 1+bc%
\end{pmatrix}%
\begin{pmatrix}
0 & -b \\ 
b^{-1} & 0%
\end{pmatrix}%
.
\end{equation*}

Let us choose a set of elements of the form (\ref{form_eq}) representing
each torus in $\mathcal{T}$ exactly once and denote this set of
representative elements by $R$. $.$

\paragraph{Generators}

The group $A$ is a cyclic group and we can find a generator $g_{A}$ for $A$.
This task is simple from the computational perspective, since the group $A$
is finite, consisting of $p-1$ elements.

Now, we make the following two observations. First observation is that the
oscillator basis $\mathcal{B}_{A}$ is the basis of eigenfunctions of the
operator $\rho \left( g_{A}\right) $.

The second observation is that, other bases in the oscillator system $%
\mathfrak{S}_{O}^{s}$ can be obtained from $\mathcal{B}_{A}$ by applying
elements from the set $R$. More specifically, for a torus $T$ of the form $%
T=gAg^{-1}$, $g\in R,$ we have 
\begin{equation*}
\mathcal{B}_{gAg^{-1}}=\{ \rho (g)\varphi ;\text{ }\varphi \in \mathcal{B}%
_{A}\}.
\end{equation*}

Concluding, we described the (split) oscillator system%
\begin{equation*}
\mathfrak{S}_{O}^{s}\mathcal{=\{}\rho \left( g\right) \varphi :g\in
R,\varphi \in B_{A}\}.
\end{equation*}

\paragraph{Formulas}

We are left to explain how to write explicit formulas (matrices) for the
operators $\rho \left( g\right) $, $g\in R$.

First, we recall that the group $Sp$ admits a Bruhat decomposition $Sp=B\cup
B\mathrm{w}B,$ where $B$ is the Borel subgroup consisting of upper
triangular matrices in $Sp$ and $\mathrm{w}$ denotes the Weyl element%
\begin{equation*}
\mathrm{w}=%
\begin{pmatrix}
{\small 0} & {\small 1} \\ 
-{\small 1} & {\small 0}%
\end{pmatrix}%
.
\end{equation*}

Furthermore, the Borel subgroup $B$ can be written as a product $B=AU=UA$,
where $A$ is the standard diagonal torus and $U$ is the standard unipotent
group 
\begin{equation*}
U=\left \{ 
\begin{pmatrix}
1 & 0 \\ 
u & 1%
\end{pmatrix}%
:u\in \mathbb{F}_{p}\right \} .
\end{equation*}

Therefore, we can write the Bruhat decomposition also as $Sp=UA\cup UA%
\mathrm{w}U$.

Second, we give an explicit description (which can be easily verified using
identity (\ref{Eg})) of operators in the Weil representation which are
associated with different types of elements in $Sp$. The operators are
specified up to a unitary scalar, which is enough for our needs.

\begin{itemize}
\item The standard torus $A$ acts by (normalized) scaling: An element 
\begin{equation*}
{\small a=}%
\begin{pmatrix}
{\small a} & {\small 0} \\ 
{\small 0} & {\small a}^{-1}%
\end{pmatrix}%
,
\end{equation*}%
acts by 
\begin{equation*}
S_{a}\left[ f\right] \left( t\right) =\sigma (a)f\left( a^{-1}t\right) ,
\end{equation*}%
where $\sigma :\mathbb{F}_{p}^{\times }\rightarrow \{ \pm 1\}$ is the
Legendre character, $\sigma (a)=$ $a^{\frac{p-1}{2}}(\func{mod}p)$.

\item The subgroup of strictly lower diagonal elements $U\subset Sp$ acts by
quadratic exponents (chirps): An element 
\begin{equation*}
u=%
\begin{pmatrix}
1 & 0 \\ 
u & 1%
\end{pmatrix}%
,
\end{equation*}%
acts by 
\begin{equation*}
M_{u}\left[ f\right] \left( t\right) =\psi (-\tfrac{u}{2}t^{2})f\left(
t\right) .
\end{equation*}%
where $\psi :\mathbb{F}_{p}\rightarrow 
\mathbb{C}
^{\times }$ \ is the character $\psi (t)=e^{\frac{2\pi i}{p}t}.$

\item The Weyl element 
\begin{equation*}
\mathrm{w}=%
\begin{pmatrix}
0 & 1 \\ 
-1 & 0%
\end{pmatrix}%
,
\end{equation*}%
acts by discrete Fourier transform 
\begin{equation*}
F\left[ f\right] \left( w\right) =\frac{1}{\sqrt{p}}\sum \limits_{t\in 
\mathbb{F}_{p}}\psi \left( wt\right) f\left( t\right) .
\end{equation*}%
Using the Bruhat decomposition we conclude that every operator $\rho \left(
g\right) $, $g\in Sp$, can be written either in the form $\rho \left(
g\right) =M_{u}\circ S_{a}$ or in the form $\rho \left( g\right)
=M_{u_{2}}\circ S_{a}\circ F\circ M_{u_{1}}$, where $M_{u},S_{a}$ and $F$
are the explicit operators above.
\end{itemize}

\begin{example}
For $g\in R$, with $b\neq 0$, the Bruhat decomposition of $g$ is given
explicitly by 
\begin{equation*}
g=%
\begin{pmatrix}
1 & 0 \\ 
\frac{1+bc}{b} & 1%
\end{pmatrix}%
\begin{pmatrix}
b & 0 \\ 
0 & b^{-1}%
\end{pmatrix}%
\begin{pmatrix}
0 & 1 \\ 
-1 & 0%
\end{pmatrix}%
\begin{pmatrix}
1 & 0 \\ 
b^{-1} & 1%
\end{pmatrix}%
,
\end{equation*}%
and consequently 
\begin{equation*}
\rho \left( g\right) =M_{\frac{1+bc}{b}}\circ S_{b}\circ F\circ M_{b^{-1}}.
\end{equation*}%
For $g\in R$, with $c=0,$ we have%
\begin{equation*}
g=%
\begin{pmatrix}
1 & 0 \\ 
u & 1%
\end{pmatrix}%
,
\end{equation*}%
and 
\begin{equation*}
\rho \left( g\right) =M_{u}.
\end{equation*}
\end{example}

\subsection{Pseudocode}

Below, is given a pseudo-code description of the construction of the (split) 
\textit{oscillator system }$\mathfrak{S}_{O}^{s}.$

\begin{enumerate}
\item Choose a prime\textrm{\ }$p.$

\item Compute generator\textrm{\ }$g_{A}$ for the standard torus $A$.

\item Diagonalize\textrm{\ }$\rho \left( g_{A}\right) $ and obtain the basis
of eigenfunctions $\mathcal{B}_{A}.$

\item For every\textrm{\ }$g\in R$:

\item Compute the operator\textrm{\ }$\rho \left( g\right) $ as follows:

\begin{enumerate}
\item Calculate the Bruhat decomposition of\textrm{\ }$g$, namely, write $g$
in the form $g=u_{2}\cdot a\cdot \mathrm{w}\cdot u_{1}$ or $g=u\cdot a$.

\item Calculate the operator\textrm{\ }$\rho \left( g\right) $, namely, take 
$\rho \left( g\right) =M_{u_{2}}\circ S_{a}\circ F\circ M_{u_{1}}$ or $\rho
\left( g\right) =M_{u}\circ S_{a}$.
\end{enumerate}

\item Compute the vectors $\rho (g)\varphi $, for every $\varphi \in B_{A}\ $%
and obtain the system $B_{gAg^{-1}}$.
\end{enumerate}

\begin{remark}[Running time]
It is easy to verify that the time complexity of the algorithm presented
above is $O(p^{4}\log p)$. This is, in fact, an optimal time complexity,
since already to specify $p^{3}$ vectors, each of length $p$, requires $%
p^{4} $ operations.
\end{remark}

{\Large Acknowledgement.}{\LARGE \ }It is a pleasure to thank J. Bernstein
for his interest and guidance in the mathematical aspects of this work. We
are grateful to S. Golomb and G. Gong for their interest in this project. We
thank B. Sturmfels for encouraging us to proceed in this line of research.
The first author would like to thank V. Anantharam, A. Gr\"{u}nbaum and A.
Sahai for interesting discussions. Finally, the second author is indebted to
B. Porat for so many discussions where each tried to understand the cryptic
terminology of the other.

\end{document}